\documentclass[12pt]{article}
\input epsf.sty

\newcommand\pcite[1]{\protect{\cite{#1}}}

\newcommand{\la}[1]{\label{#1}}
 
\newcommand{\be}{\begin{equation}}
\newcommand{\ee}{\end{equation}}
\newcommand{\ba}{\begin{eqnarray}}
\newcommand{\ea}{\end{eqnarray}}
\newcommand{\bi}{\begin{itemize}}
\newcommand{\ei}{\end{itemize}}
\newcommand{\nr}[1]{(\ref{#1})}

\newcommand{\fr}[2]{{\frac{#1}{#2}}}

\newcommand{\<}{\langle} 
\renewcommand{\>}{\rangle}  

\newcommand{\ev}[1]{\langle #1 \>}

\newcommand{\eq}{Eq.~}
\newcommand{\eqs}{Eqs.~}
\newcommand{\fig}{Fig.~}
\newcommand{\figs}{Figs.~}

\newcommand{\op}{\theta}
\newcommand{\avphi}{\phi^2_{\rm av}}
\newcommand{\avphione}{\phi_{\rm av}}
\newcommand{\avphiC}{\phi^2_{{\rm av},C}}
\newcommand{\avphiX}[1]{\phi^2_{{\rm av},#1}}
\newcommand{\lat}{{\ell}}
\newcommand{\bd}{{\bf d}}
\newcommand{\dpdt}{\langle | \Delta\avphi/\Delta t|\>}

\newcommand{\tx}[1]{\textrm{#1}}
\newcommand{\h}{\hspace*{1cm}} 
\newcommand{\lf}{\lambda_1/\lambda_2}

\def\lsi{\raise0.3ex\hbox{$<$\kern-0.75em\raise-1.1ex\hbox{$\sim$}}}
\def\gsi{\raise0.3ex\hbox{$>$\kern-0.75em\raise-1.1ex\hbox{$\sim$}}}
\newcommand{\lsim}{\mathop{\lsi}}
\newcommand{\gsim}{\mathop{\gsi}}
\makeatletter \@addtoreset{equation}{section} \makeatother

\makeatletter
\renewcommand\section{\@startsection {section}{1}{\z@}%
                                   {-3.5ex \@plus -1ex \@minus -.2ex}%
                                   {2.3ex \@plus.2ex}%
                                   {\normalfont\large\bfseries}}
\renewcommand\subsection{\@startsection{subsection}{2}{\z@}%
                                     {-3.25ex\@plus -1ex \@minus -.2ex}%
                                     {1.5ex \@plus .2ex}%
                                     {\normalfont\normalsize\bfseries}}
\renewcommand\thesection {\@arabic\c@section}
\renewcommand\thesubsection   {\thesection.\@arabic\c@subsection}
\makeatother

\begin{document}
 
\begin{titlepage}
\begin{flushright}
NORDITA-2001/6 HE\\
UW-PT 01-05\\
hep-lat/0103036\\
\end{flushright}
\begin{centering}
\vfill
 
{\bf Non-perturbative computation of the bubble nucleation rate in the
     cubic anisotropy model}

\vspace{0.8cm}
 
Guy D. Moore$^{\rm a,}$\footnote{guymoore@phys.washington.edu},
Kari Rummukainen$^{\rm b,d,}$\footnote{kari@nordita.dk}
and
Anders Tranberg$^{\rm c,e,}$\footnote{anderst@science.uva.nl}

\vspace{0.3cm}
{\em $^{\rm a}$Department of Physics, University of Washington,
Seattle, WA 98195-1560 USA\\%
}
\vspace{0.3cm}
{\em $^{\rm b}$NORDITA and $^{\rm c}$Niels Bohr Institute\\
 Blegdamsvej 17, DK-2100 Copenhagen \O, Denmark\\%
}
\vspace{0.3cm}
{\em $^{\rm d}$Helsinki Institute of Physics\\
P.O.Box 64, FIN-00014 University of Helsinki, Finland\\%
}
\vspace{0.3cm}
{\em $^{\rm e}$Institute for Theoretical Physics, University of
Amsterdam \\ 
Valckenierstraat 65, 1018 XE Amsterdam, The Netherlands\\%
}

\vspace*{0.7cm}
 
\end{centering}
 
\noindent
At first order phase transitions the transition proceeds through
droplet nucleation and growth.  We discuss a lattice method for
calculating the droplet nucleation rate, including the complete
dynamical factors.  The method is especially suitable for very
strongly suppressed droplet nucleation, which is often the case in
physically interesting transitions.  We apply the method to the
3-dimensional cubic anisotropy model in a parameter range where the
model has a radiatively induced strong first order phase transition,
and compare the results with analytical approaches.
\vfill
\noindent
 

\vspace*{1cm}

\vfill
 
\noindent
March 2001  

\end{titlepage}
 

\section{Introduction}

Many first order phase transitions in Nature happen so that the external
parameter which drives the transition varies on timescales several
orders of magnitude longer than the microscopic interaction timescale.
Examples include freezing or vaporization of liquids, several
magnetization transitions, and a number of phase transitions in the
early Universe.  Despite the very slow change in the control
parameter, these transitions do not happen immediately when the
thermodynamic transition point of the control parameter is reached,
but the system can remain in the old, metastable phase for an
extremely long time compared to the interaction timescales.
(For the case where a temperature change induces the phase transition,
this is called supercooling or superheating.)  In
homogeneous systems, the transition happens through spontaneous
nucleation of droplets (or bubbles) larger than a critical size, and
subsequent droplet growth and coalescence.

Let us, for a moment, specialize to temperature driven phase
transitions.  Given a fixed cooling rate, the depth of supercooling is
determined by the temperature dependent nucleation rate of critical
droplets.  The more supercooling, the farther the system is driven
away from thermodynamical equilibrium, and when the transition
finally happens, the more dramatic the consequences will be:
hydrodynamic flows, shock waves, entropy production, generation of new
length scales etc.  In order to make quantitative predictions
we need to know the nucleation rate accurately.\footnote{
As an example, the electroweak phase transition in
the early Universe may have been of first order, in which case
the dominance of matter over antimatter may arise as a result of
strong metastability.  However, the efficacy of the generation
mechanism depends sensitively on the quantitative details of the
transition and the degree of the supercooling.
For a review, see, for example \cite{RubakovShaposhnikov}.}

A statistical theory of droplet nucleation in liquids was formulated
already in 1935 by Becker and D\"oring~\cite{BeckerDoring}.  Later,
Cahn and Hilliard~\cite{CahnHilliard} gave a phenomenological
description of the nucleation in Ginzburg-Landau theory.  Finally, the
theory of nucleation was given a firm field theory foundation by
Langer~\cite{Langer} (for a review, see Ref.~\cite{GuntonDroz}).
Langer's method was generalized to relativistic quantum field theories
by Callan and Coleman~\cite{CallanColeman} and
Voloshin~\cite{Voloshin}, and extended to finite temperature quantum
field theories by Affleck~\cite{Affleck} and Linde~\cite{Linde}.

A fully consistent application of Langer's theory can be difficult.
This is especially so in theories with a {\em radiatively induced}
first order transition.  The tree-level Lagrangian of these theories
does not show a first order transition; the transition becomes
apparent --- the effective potential develops more than one minimum
--- only after some field fluctuations have been integrated over.  On
the other hand, in nucleation theory we start from a potential which has
two minima, one of them metastable, and find the classical droplet
solution of the equations of motion, with energy $H$.  The
nucleation rate (per time and volume) is now proportional to the
Boltzmann factor: $\Gamma = D\exp(-H/T)$.  Langer's theory gives us a
recipe how to evaluate the constant of proportionality $D$; an
essential ingredient is the determinant of the fluctuations around the
classical critical droplet solution.  The problem with the radiatively
induced transitions is now apparent: we already ``used up'' some
fluctuations for evaluating the effective potential.  Thus, care has
to be taken to ensure the proper counting of the fluctuations.  Note
also that the fluctuation determinant represents only a one loop
treatment of fluctuations and ignores nonlinearities between
fluctuations; in some cases these are very important, and a high loop or
nonperturbative treatment is needed.

In this paper we study the nucleation rate in the {\em cubic anisotropy
model\,}, a theory with two scalar fields, using numerical lattice
simulations.  It is one of the simplest field theories with a
radiatively generated first order phase transition, and thus well
suited for both analytical and numerical studies.  Partial results
have already been published in Ref.~\cite{SEWMposter}.  The phase
structure of the theory has been studied using $\epsilon$ expansion
techniques \cite{Rudnick,AYcubic} and Monte Carlo simulations
\cite{ASYZcubic}.  The droplet nucleation rate itself has been studied
using coarse grained effective potentials \cite{StrumiaTetradis}.

In principle, it is relatively straightforward to study the nucleation
rate using standard real-time lattice simulations, provided that the
metastability is not strong: one simply prepares initial
configurations in the metastable phase, lets them evolve in time and
waits for the transitions to happen.  This method has been used in
simulations of kinetic discrete spin models \cite{Ising3,Arkin} and in scalar field
theories \cite{Borsanyi}.  However, it is applicable only if
the metastability time scale is, at most, a few orders of magnitude
longer than the microscopic interaction time scale.  Otherwise the
system simply remains metastable for the duration of any practical
computer simulation.  This is the situation for many phase transitions
in Nature.

In contrast, the method described in this paper is applicable to first
order transitions with almost arbitrarily strong metastability.\footnote{
We note that algorithms exist in the literature for studying
metastable states in some {\em discrete} spin models (for example, the
Ising model), and for specific choices for update dynamics.  For a
review, see Ref.~\cite{Novotny} and references therein.}
Our method is by no means specific to the cubic anisotropy model; it can be
used to study the nucleation rate in any theory, provided that:

(1) The thermodynamics of the system is well described by 
an (essentially) classical
partition function, and 
both the thermodynamics and the real time evolution of the
system can be reliably simulated on the lattice;

(2) there is an ``order parameter like'' observable which can 
resolve the phases and the potential
barrier between them sufficiently accurately; and

(3) the barrier is large and the nucleation rate is small.

In a simultaneous calculation this method was applied to SU(2) gauge +
Higgs theory, which is an effective theory for the electroweak phase
transition \cite{ewbubble}.  With suitable generalizations one can
apply the method also for almost any strongly suppressed process in
various theories on and off the lattice.  Indeed, the method here is
closely related to the one used for studying the ``broken phase
sphaleron rate'' in the electroweak theory \cite{broken_nonpert}.

The Monte Carlo simulations are free of the ambiguities and problems
which analytical calculations suffer from in practice.  Relatively
modest computational resources give us the nucleation rate, as a
function of the supercooling, quite accurately --- the calculations in
this work were performed using only standard workstations and PC's.
We compare the results with various (semi)analytical 
approaches: the thin wall approximation, where we use latent heat and
surface tension determined by lattice simulations; purely
perturbative analysis; and the coarse grained effective potential
calculation by Strumia and Tetradis \cite{StrumiaTetradis}.  The
results from these calculations show huge sensitivity 
to various approximations in analytical calculations.
On the other hand, we do not observe any signs of the 
breakdown of Langer's theory of homogeneous nucleation in radiatively
generated transitions, reported in Ref.~\cite{StrumiaTetradis}.

This paper is structured as follows: in Sec.~\ref{sec:howto} we give a
detailed description of our lattice approach.  Since the method is
quite general, the discussion is completely independent of the
specifics of the model studied and can be read independently of the
rest of the paper.  This section is a reformulation and development of
the discussion in Sec.~{II} in Ref.~\cite{ewbubble}.  The cubic
anisotropy model, its lattice discretization, and its real time
evolution is presented in Sec.~\ref{sec:cubic}.
The thermodynamics of the model are studied nonperturbatively on the
lattice in Sec.~\ref{sec:numerics}, 
and the nucleation rate is determined in
Sec.~\ref{sec:rate}.  The comparison with analytical and
semianalytical approaches is in Sec.~\ref{sec:compare}, and finally, we
conclude in Sec.~\ref{sec:conclusion}.

\section{How to calculate the droplet nucleation rate}\la{sec:howto}
\newcommand{\vfrac}{X}

In this section we describe the general framework of the method for
calculating the droplet nucleation rate with lattice Monte Carlo
methods.  The calculation can be split into two parts: (1) the
measurement of the {\em probability} of the critical droplets, and (2)
the calculation of the {\em dynamical} factors which convert the
probability into a rate.  In the case we are interested in, very weak
supercooling, the probabilistic suppression is by far the dominant
part.  Thus, for many purposes, a sufficiently accurate answer can be
obtained by calculating the droplet probability in the way described
below, and multiplying it with suitably chosen dimensional factors in
order to convert the probability into a rate.  However, without the
correctly calculated dynamical information the resulting rate can be
off by several orders of magnitude.  Moreover, only by including the
dynamical information does the result become truly independent of the
choice of order parameter and other details of the procedure.

In order to make the discussion more concrete, we shall adopt the
terminology from the cubic anisotropy model (described in
Sec.~\ref{sec:cubic}), and consider a first order symmetry breaking
transition from a {\em symmetric phase} to a {\em broken phase},
which occurs when a control parameter $m^2$ is lowered below the
(adiabatic, infinite volume)
transition value $m^2_c$.
However, we emphasize that the method
described here is not specific to the cubic anisotropy model or even
to order-disorder transitions in general; indeed, by suitable
generalization it can be applied to any system where we have two or
more states separated by a large potential barrier, and where the
system evolves in phase space with some dynamical prescription
amenable to computer simulations.

\subsection{The probability of a critical droplet}\la{sec:droplet}

Let us consider a system in a finite cubical volume $V=L^3$, with
periodic boundary conditions.  At the transition point $m^2 = m^2_c$
the probability distribution $P(\op)$ of a suitably chosen volume
averaged ``order parameter''\footnote{What we mean here by the ``order
parameter'' is a quantity which has clearly different expectation
values in the two phases, but by no means do we require that it is a
true mathematical order parameter in the sense that it would have a
zero expectation value in one phase and a non-zero value in the other.
Thus, the transition here can be one without global symmetry 
breaking (for example, a liquid-gas transition where the density
can act as an order parameter). 
The choice of the order parameter is discussed in more detail
in Sec.~\ref{sec:orderp}.}
\be
   \op = \fr1{V} \int d^3 x \op(x),
\ee
attains the familiar two-peak form, see \fig\ref{fig:twopeak}.  The
two peaks correspond to the symmetric and broken phases, with order
parameter expectation values $\op_{\rm symm.}$ and $\op_{\rm broken}$,
respectively.  The intermediate region between the peaks consists of
configurations with co-existing domains of the symmetric and the
broken phases.  We assume here that the system size is much
larger than the thickness of the phase interfaces.  The probability
distribution can also be written in terms of the {\em constrained
free energy\,}: $P(\op) \propto \exp[ -F(\op)]$.  (Here and throughout
we rescale free energies $F$ and energies $H$ by $T$ so they are
dimensionless.)

The shape of $P(\op)$ can be readily understood by 
considering the following ``thin-wall'' mean-field
approximation: the order parameter in the bulk symmetric and broken
phases is fixed to respective expectation values, and the phase
interfaces have zero thickness, with a surface tension $\sigma$.  The
free energy densities of the symmetric and broken phases are equal,
whereas the mixed phase has an additional free energy contribution,
which equals $\sigma$ times the area of the phase interfaces. The
geometry of the mixed symmetric and broken phase domains now settles
in configurations which minimize the surface area.  In this case the
value of the normalized order parameter $\vfrac = (\op - \op_{\rm
symm.})/(\op_{\rm broken} - \op_{\rm symm.}) $ exactly equals
the volume fraction filled by the broken phase.

Given a value of $\vfrac$, it is a straightforward exercise to find
the mixed phase geometry which minimizes the area.  When $\vfrac$ is
small, we have a small, spherical droplet of the broken phase embedded
in the bulk of the symmetric phase.  However, when the diameter of the
droplet approaches the length of the box $L$, it becomes advantageous
for the broken phase to form a cylinder spanning the box along one
direction --- because of the periodic boundary conditions, the
cylinder does not have any ``end caps,'' and its area is smaller for a
given volume.  Analogously, when $\vfrac$ is further increased, the
broken phase forms a slab which spans the volume along two directions.
When $\vfrac>1/2$, the same sequence occurs again, with the symmetric
and broken phases interchanged.  Quantitatively, the stability ranges
for different mixed phase geometries, in the thin wall approximation,
are as follows: 

\noindent
\begin{tabular}{lll}
(1) droplet:  &$0 \: \: < \vfrac \le \fr{4\pi}{81}$,        
	&  area $= (36\pi \vfrac^2)^{1/3} L^2$ \\
(2) cylinder: &$\fr{4\pi}{81} < \vfrac \le \fr1{\pi}$, 
	&  area $= 2(\pi \vfrac)^{1/2} L^2$ \\
(3) slab:     &$\fr{1}{\pi}\; < \vfrac \le 1-\fr{1}{\pi}$,  
	&  area $= 2 L^2$ \, .\\
\end{tabular}

\begin{figure}[t]
\centerline{
\epsfxsize=7cm\epsfbox{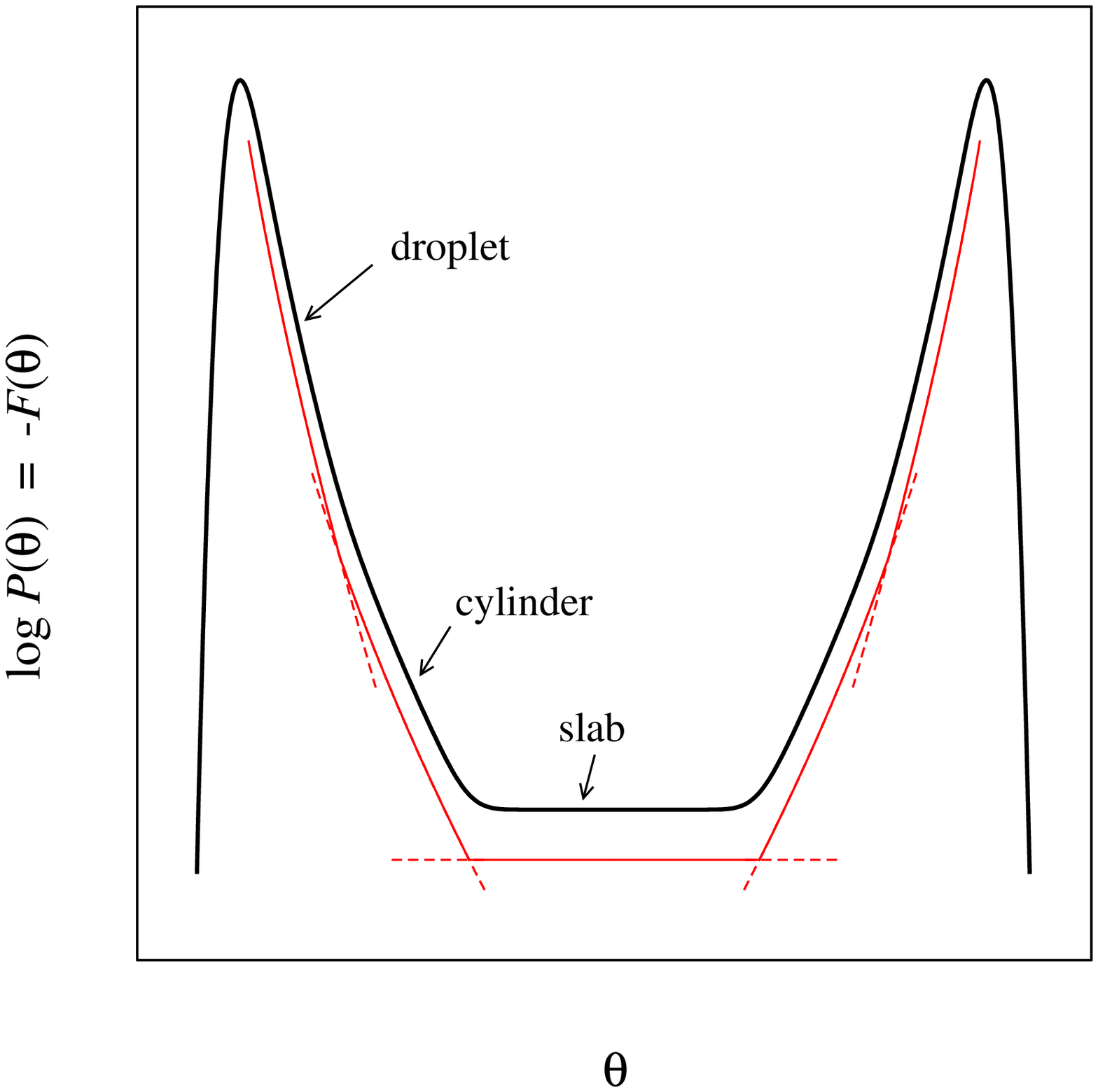} ~~~ 
\raisebox{1.1cm}{\epsfxsize=7.5cm\epsfbox{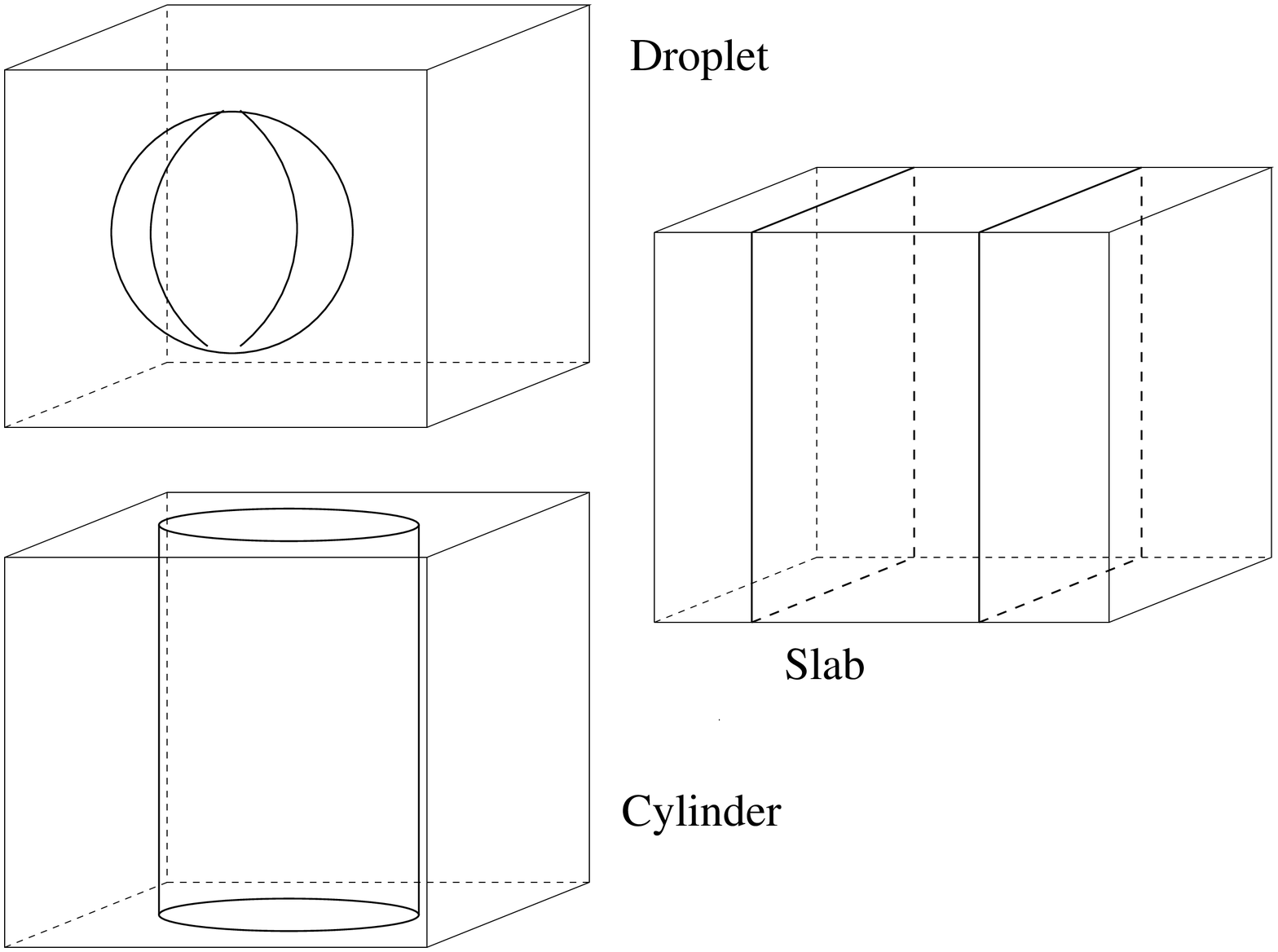}}
}
\caption[a]{Order parameter probability
distribution $P(\op)$ at the transition point $m_c^2$ in a cubic
box with periodic boundary conditions.  Thick line: A schematic plot
of the distribution on a large but finite volume.  Thin line: ``thin
wall'' approximation, consisting of droplet, cylinder and slab
topologies.}
\la{fig:twopeak}
\end{figure}

\noindent
These are shown with the thin lines in \fig\ref{fig:twopeak}.  Note
that the change in geometry is by no means continuous: for instance,
the change from a droplet into a cylinder at $\vfrac=4\pi/81$ must
occur via a deformation which increases the area substantially.  This
causes an energy barrier between different mixed phase topologies, and
the topology sectors have overlapping metastability
branches.%
\footnote{Indeed, multicanonical Monte Carlo simulations (to
be discussed in Sec.~\ref{sec:multi}), which, by construction,
overcome the huge probability suppression of the mixed phase, still
suffer from the milder tunneling barriers associated with the changes
of topology.  This is because the order parameter remains
constant when the topology change occurs.  Thus, in multicanonical
simulations at large volumes there is a clear reluctance for the
system to go from the droplet branch into the slab branch, for
example.}

What happens to the probability distribution when we lower $m^2$
slightly from the transition value $m_c^2$?  Now the free energy
densities in the symmetric and broken phases are different:
\be
  \Delta f = f_{\rm symm.} - f_{\rm broken} \simeq \delta m^2 \times 
        \left[ \fr{d f_{\rm symm.}}{d m^2} - \fr{d f_{\rm broken}}{d m^2}
        \right]_{m^2 = m_c^2} = \delta m^2 \lat\,,
\ee
where $\delta m^2 = m_c^2 - m^2$ and $\lat$ is 
proportional to the ``latent heat'' of
the transition (remember that $m^2$ is the temperature parameter).
Remaining in our simple thin wall approximation, and further assuming
that $\sigma$ remains constant as we change $m^2$, the free energy of
the mixed phase configurations has an additional contribution
proportional to the volume:
\be
  F(\vfrac,m^2) = 
  F(\vfrac,m_c^2) + \delta m^2 \lat \vfrac L^3 + \mbox{const.}
\ee
In particular, the free energy of a broken phase spherical droplet of
volume $V_d$ and area $A_d$ becomes
\be
 F_{\rm droplet}(\vfrac,m^2) - F_{\rm symm.}
	= -\delta m^2 \lat V_d + \sigma A_d
	= -\delta m^2 \lat \vfrac L^3 
	+ \sigma (36\pi)^{1/3} \vfrac^{2/3} L^2 \, .
\ee
The droplet free energy has a maximum $F_c = 16 \pi\sigma^3/(3 (\delta
m^2)^2 \lat^2)$ at droplet radius $R_d = 2\sigma/(-\delta m^2 \lat)$.
This is the critical droplet size and free energy at supercooling
$\delta m^2$: droplets smaller than this size can continuously reduce
their free energy by shrinking, and larger droplets can by growing.  The
droplet free energy is shown schematically in \fig\ref{fig:droplet}.
In a finite box the droplet volume should not exceed the droplet
$\leftrightarrow$ cylinder stability limit $L^3 4
\pi/81 \approx 0.155 L^3$, as otherwise the critical ``droplet'' is a
cylinder and the box size is essential to determining its behavior,
which is unphysical if we are interested in the behavior in the
thermodynamic limit.  Applied to the critical droplet, this 
implies that the inequality $\delta m^2 \lat > 6\sigma/L$ should be
satisfied.

\begin{figure}[t]
\centerline{\epsfxsize=8cm\epsfbox{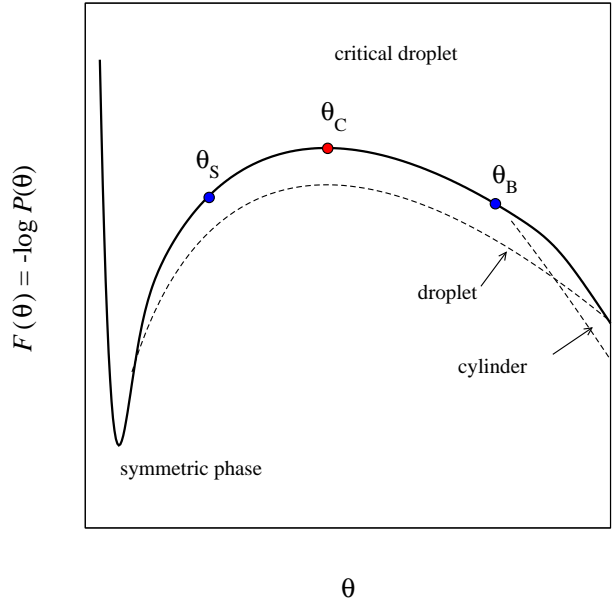}}

\caption[a]{The constrained free energy $F(\vfrac) = -\log P(\vfrac)$ 
at a small supercooling $\delta m^2$.}
\la{fig:droplet}
\end{figure}

Naturally, in a more realistic case the situation is not as simple as
in the thin wall case.  The phase interfaces have a finite thickness,
the shape of the mixed phase domains fluctuates, as also does the
value of the order parameter, averaged over the volumes of the pure
phase domains.  Nevertheless, provided that the volume is sufficiently
large, the basic topological structure outlined above remains --- even
if the shape of the cylinders, droplets and slabs becomes distorted,
the magnitude of the fluctuations is restricted by the increase in the
free energy (area$\times \sigma$) of the domains.\footnote{In order
to be able to properly define the domain shape in fluctuating, real
systems, one should {\em coarse-grain} the order parameter up to, at
least, the length scale given by the bulk correlation length.  In the
calculations in this paper we use {\em only} volume averages of the
order parameter, and the coarse-graining is of no consequence.}  The
resulting probability distribution is a ``smeared'' version of the
thin-wall case, with the pure phase $\delta$-peaks rounded into
Gaussian distributions.  This is shown as thick lines in
\figs\ref{fig:twopeak} and \ref{fig:droplet}.

The probability distribution $P(\op)$ is to be measured with Monte
Carlo simulations.  Due to the enormous suppression of the critical
droplets, standard Monte Carlo simulations, where configurations are
sampled with weight $p \propto \exp[-H]$, are completely inadequate
for the task.  However, the {\em multicanonical\,} Monte Carlo method
\cite{BergNeuhaus} has been developed for just this kind of task; this
will discussed in Sec.~\ref{sec:multi}.  From the
measured probability distribution $P(\op) \propto
\exp(-F(\op))$ at small supercooling $\delta m^2$
we can
measure the critical droplet free energy  
\be
  F_C(\delta m^2) \approx F(\op_C) - F(\op_{\rm symm.}) = 
	- \log \fr{P(\op_C)}{P(\op_{\rm symm.})}\,, \la{fdroplet}
\ee
where $\op_C$ is the local maximum of the free energy, and $\op_{\rm
symm.}$ is the location of the symmetric phase minimum. 
Since one should remain a safe
distance away from the cylinder and slab regions, $\vfrac$ should remain
well below $\approx 0.15$.
This limits the range of $\delta m^2$ which can be probed
at a given $L$, or sets what $L$ is required for a given $\delta m^2$.
We note that \eq\nr{fdroplet} resembles the widely used
histogram method for calculating the planar interface surface tension
$\sigma$, proposed by Binder \cite{Binder}.  In some cases
\eq\nr{fdroplet} can already give a fair estimate of the droplet
nucleation rate:
\be
  \Gamma \propto \mbox{[time]}^{-1}\mbox{[length]}^{-3} \exp(-F_C)\,,
\la{naivegamma}
\ee
where [time] and [length] are suitably chosen dimensionful constants.
Unfortunately,
\eq\nr{naivegamma} tells us nothing about the constant of proportionality,
and the dynamical information is missing completely.  The result also
explicitly depends on our choice of the order parameter.  Thus, the
rate can be off by orders of magnitude.  Furthermore, as opposed to
Binder's method, which gives 
the surface tension $\sigma$ exactly as
$V\rightarrow\infty$, \eq\nr{fdroplet} does not have a well-behaved
infinite volume limit with fixed $\delta m^2$.  The main problem is
that, for a fixed supercooling and droplet size, we have
$\op_C - \op_{\rm symm} \propto 1/V$, but the width of the
Gaussian-like symmetric phase peak is $\propto 1/V^{1/2}$.  Thus, the
symmetric phase fluctuations will ``swallow'' $\op_C$ when the 
volume is too large!  In practice, this does not appear to be a
problem for suitably chosen order parameters (see Sec.~\ref{sec:orderp}).

\subsection{Rate of droplet nucleation}\la{subsec:rate}

Although the droplet nucleation rate is inherently a
non-equilibrium problem, in a finite (but large) volume it can be
formulated fully in thermal equilibrium.  Let us take an ensemble of
configurations in thermal equilibrium at small supercooling $\delta
m^2$, and let each of the configurations evolve according to some
dynamical prescription.  Naturally, we require that the evolution
preserves the equilibrium ensemble, and we consider here only 
Markovian dynamics which satisfy {\em detailed balance:}
\be
  \fr{T[{(\phi_a,\pi_a) \rightarrow (\phi_b,\pi_b)}]} 
     {T[{(\phi_b,{-\pi_b}) \rightarrow (\phi_a,{-\pi_a})}]} = 
    \exp[H(\phi_b,\pi_b)-H(\phi_a,\pi_a)] \,.
\la{balance}
\ee
Here $\phi$ are field variables, $\pi$ the momentum variables
(which may not exist in the dynamics at all).  $H(\phi,\pi) =
H(\phi,-\pi)$ is the energy of the state (scaled by $T$ to be
dimensionless), and $T(a\rightarrow b)$ is
the transition probability (along some trajectory) from state $a$ to
$b$.  All standard dynamics we know of obey this condition
(for example, Hamiltonian, Langevin, Glauber, heat bath).

We specialize to supercooling,\footnote%
{The case of superheating, tunneling out of the broken phase, is a
trivial extension.}
so $m^2 < m_c^2$, and at any small time interval the vast majority of
the configurations remain in the broken phase, a much smaller fraction
are in the symmetric phase, and a tiny number of configurations are in
the process of evolving from one phase into another.  In equilibrium
the rate must be equal in both directions.
Let us choose two order parameter values $\op_S$ and $\op_B$ on the
symmetric and broken phase sides of $\op_C$, chosen so that the
constrained free energy $F(\op_S),\, F(\op_B) \ll F(\op_C)$; see
\fig\ref{fig:droplet}.  If we now take a configuration from the thermal
distribution with order parameter restricted to $\op_S$ and let it
evolve in time, due to the steep slope in the free energy, it is
almost certain {\em not} to evolve to $\op_C$ in the ``medium term,''
i.e.~in timescales shorter than the very long tunneling time $1/\Gamma
V$.  Similarly, a configuration at $\op_B$ is overwhelmingly more
likely to fall down into the broken phase than to go to the critical
droplet.

During the tunneling the configuration must evolve along a continuous
trajectory through the order parameter values
$\op_S\rightarrow\op_C\rightarrow\op_B$.  We shall focus on the
piece of the real time evolution trajectory between
the {\em last} time the order parameter had the value $\op_S$,
before reaching $\op_B$ for the {\em first} time.  Since the time
evolution of the order parameter may be ``noisy,'' this trajectory
may cross the value $\op_C$ several times in a short time
interval.  We note here that the values of $\op_S$ and $\op_B$ are not
critical, as long as the free energy condition $F(\op_{S,B}) \ll
F(\op_C)$ is satisfied.  For example, we could choose $\op_S$ to be
well within the bulk of the symmetric phase; the point is that the
trajectory $\op_S\rightarrow\op_B$ can be used to identify the
tunneling events.

We define the symmetric phase probability $P_{\rm symm.}$ to be the
integral of the canonical distribution $P(\op) \propto
\log[-F(\op)]$ up to up to the critical droplet value $\op_C$,
\be 
  P_{\rm symm.} = \int_{\theta_{\rm min}}^{\op_C} d\op P(\op)\, ,
\ee
with $\theta_{\rm min}$ the smallest possible value of $\theta$.
The choice of $\op_C$ here is for convenience; since by far the
dominant contribution to the integral comes from the
central peak of the symmetric phase, choosing some other
reasonable value in the neighborhood of $\op_C$ would give
only exponentially suppressed corrections.

Let $n_s(t)$ be the probability that a configuration which was in
the symmetric phase at time $t=0$ still remains there at time $t$,
normalized so that $n_s(t=0) = P_s$.  The tunneling rate 
$\Gamma$ is defined through
\be
 \Gamma V n_s(t) = -\fr{d n_s(t)}{d t} \h \rightarrow \h
  n_s(t) = n_s(0) \exp\left[ -\Gamma V t \right]\,.
  \la{exprate}
\ee
This immediately suggests that we can measure $\Gamma$ by simply
measuring the average time $\tau$ it takes for the symmetric phase
configurations to tunnel: $\Gamma = 1/(V\tau)$.  However, when the
rate is very small, as it is in our case, this approach is utterly
impractical.\footnote{
Strictly speaking, at {\em extremely} long times 
\eq\nr{exprate} is not correct; it neglects the multiple tunneling trajectories
symmetric $\rightarrow$ broken $\rightarrow $ symmetric etc.  Indeed,
in the limit $t\rightarrow\infty$,  $n(t) \approx P_{\rm symm.}^2$.
However, since the rate of the the inverse process is much smaller
than $V\Gamma$, this becomes significant only at times which are much
longer than the tunneling time $(\Gamma V)^{-1}$.}


Let us now consider times which are much shorter than the tunneling time,
$t \ll 1/(V \Gamma)$ (this will not restrict the validity of the final
results).  In this case the symmetric phase consists
almost entirely of configurations which were there already at $t=0$.
We can substitute $n_s(t) \approx n_s(0) = P_s$ on the l.h.s. of
\eq\nr{exprate}, and $dn_s(t)/dt$ becomes the
time-independent flux of configurations from the symmetric to the
broken phase.  The central idea of the method described here is the
realization that, since all of the tunneling trajectories must go
through $\op_C$, we can calculate the flux $dn_s/dt$ by
sampling configurations and trajectories directly at $\op =
\op_C$.  Labelling the configurations with symbol $\phi$ and the
trajectories through $\phi$ 
with $\alpha_\phi$, the flux can be written as
\be
  \fr{dn_s}{dt} = 
  \int d\phi d\alpha_\phi \; A e^{-H(\phi,\alpha_\phi)} \; 
	\delta(\op(\phi) - \op_C) \,
        \times \fr{d \op(\alpha_\phi)}{d t}  
	\times \delta^\alpha_{S\rightarrow B}
\la{flux}
\ee
Here $A \exp{-H(\phi,\alpha_\phi)}$ gives the thermal equilibrium
probability of a configuration $\phi$ and a trajectory $\alpha_\phi$ which
goes through it.  In other words, we want to average over all
configurations $\phi$ such that 
$\op(\phi)=\op_C$, and all trajectories which go
through the configuration $\phi$, with the correct, thermal weight.
$A$ is a normalization constant, which satisfies
\be
  \int d\phi d\alpha_\phi  A e^{-H(\phi,\alpha_\phi)} 
	\delta(\op(\phi) - \op) = P(\op),
\ee
where $P(\op)$ is the order parameter probability distribution.

The dynamical information is contained in the last two terms of the
integrand in \eq\nr{flux}.  The derivative $ d \op(\alpha_\phi)/d t$ is
evaluated along the trajectory $\alpha_\phi$ at the point where
$\alpha_\phi$ crosses the pivot configuration $\phi$.  Its purpose is to
cancel the Jacobian factor arising from the $\delta$-function: along the
trajectory, 
\be
  \delta(\op-\op_C) = \left| \fr{d\op}{dt}\right|^{-1}\delta(t-t_C),
\la{jacobian}
\ee
where $t_C$ is the time when the configuration crosses the point
$\op_C$.  Thus, it turns the integrand from a probability density of
configurations at $\op_C$ to a flux of trajectories through the
value $\op_C$.  This can be understood in another way: if
$d\op/dt$ is small for trajectory $\alpha$, a configuration evolving
along it loiters near $\op_C$ longer than a configuration on a
trajectory where $d\op/dt$ is large.  Thus, a ``slow'' trajectory
contributes more to the probability density
$e^{-H}\delta(\op-\op_C)$ than a ``fast'' one; and the inclusion of
$d \op(\alpha)/d t$ exactly cancels this bias.

Finally, in order to count only trajectories which really
tunnel from the symmetric phase to the broken phase, we define
the projection
$\delta_{S\rightarrow B}^\alpha = 1$, if $\alpha$ tunnels from $\op_S$
to $\op_B$; $\delta_{S\rightarrow B}^\alpha = 0$ otherwise.  As
discussed above, we define the tunneling trajectories to be ones which
evolve from $\op_S$ through the configuration $\phi$ to $\op_B$.  The
result is that the integral in \eq\nr{flux} evaluates the {\em number
density\,} of tunneling trajectories through $\op_C$, which is exactly
the rate we are after.

During the evolution from $\op_S$ to $\op_B$ the tunneling
trajectories may cross $\op_C$ several times (always an odd number), and
these trajectories are counted by the integral \nr{flux} at every
crossing.  There is no overcounting, however: if there are $(2n{+}1)$
crossings, in $(n{+}1)$ of these $d \op/dt$ is positive, and in $(n)$ it
is negative,
corresponding to crossings left$\rightarrow$right and
right$\rightarrow$left, respectively.  The absolute magnitude of the
integrand is equal at every crossing of $\avphiC$ along the trajectory:
it follows from \eqs\nr{balance}, \nr{flux} and 
\eq\nr{jacobian} that the probability of choosing a crossing 
configuration $\phi_a$ and a trajectory which evolves to another
crossing configuration $\phi_a\rightarrow \phi_b$, is equal to
choosing one evolving $\phi_b\rightarrow \phi_a$.  Thus, $n$ positive
and negative contributions cancel, and the trajectory is counted only
once.\footnote{See also the discussion in the appendix A of
Ref.~\cite{ewbubble}.}

This also implies that the dynamical factors in \eq\nr{flux} can be 
rewritten in the following form, which is much more suitable for
numerical work:
\be
   \fr{d \op(\alpha)}{d t} \times \delta^\alpha_{S\rightarrow B} \,
   \longrightarrow 
   \fr12
   \bigg| \fr{d \op(\alpha)}{d t}\bigg | \times
	     \fr{\delta^\alpha_{\rm tunnel}}{N^\alpha_{\rm crossings}}\,.
\la{flux2}
\ee
Here $N^\alpha_{\rm crossing}$ is the number of times the trajectory
$\alpha$ crosses $\op_C$, and $\delta^\alpha_{\rm tunnel} = 1$ for
trajectories which tunnel {\em either $\op_S\rightarrow \op_B$ or
$\op_B\rightarrow \op_S$}: since the whole ensemble is in thermal
equilibrium, the flux of trajectories from the symmetric to the broken
phase must equal the inverse one.  We can take an average over the two
sets of trajectories, hence the factor $1/2$ in \nr{flux2}.  Using the
expression \nr{flux2} makes the integrand in \eq\nr{flux} positive
definite, which improves the convergence of the numerical
Monte Carlo integration.

In \eq\nr{flux} we implicitly assumed that the trajectory
is differentiable in $t$.  This is true for Hamiltonian type
dynamics, but not for ``noisy'' time evolution, like
Langevin or Glauber dynamics.  In these cases we must
substitute the derivative with a finite difference:
\be
  \fr{d\op}{dt} \rightarrow \fr{\Delta\op}{\Delta t}\,,
\ee
and the trajectory must be sampled with the same frequency $\Delta t$ when
the crossings in \eq\nr{flux2} are counted.  Due to the noisiness of
$\op(t)$, the smaller $\Delta t$ is, the more crossings of $\op_C$ are
potentially resolved.  However, this is compensated by the increase in
$\Delta\op/\Delta t$.  For example, in Langevin dynamics,
when $\Delta t$ is small enough,
the trajectory $\op(t)$ behaves almost like a Brownian random walk in
the proximity of $\op_C$.  In this case $\< |\Delta \op/ \Delta
t| \> \sim (\Delta t)^{-1/2}$, and $\< N_{\rm
crossings}\> \sim (\Delta t)^{-1/2}$.  Thus, neither of these
terms have a $\Delta t\rightarrow 0$ limit, but the product in
\eq\nr{flux2} has.

The use of finite time differences arises automatically in numerical
time evolution, also in Hamiltonian evolution, and we shall use this
formulation from now on.  It is natural to use the same $\Delta t$ in
sampling the trajectory as is the step size of the time evolution;
however, this is not necessary and the sampling interval can be
longer.  However, it should not be so long that the free energy
$F(\op)$ changes appreciably in one interval.  In practice this is not
a problem.

\subsection{Monte Carlo evaluation}\la{sec:mce}

In practical Monte Carlo simulations we evaluate \eq\nr{flux} as
follows: let us select configurations from the thermal ensemble,
$p\propto e^{-H}$, but with the order parameter restricted to a
narrow range $|\op - \op_C| < \epsilon/2$.  We can now construct
trajectories (one or more for each configuration) going through $\phi$
by evolving the configuration {\em both forward and backwards\,} in
time, until both the forward and the backwards ends of the trajectory
reach order parameter values $\op_S$ or $\op_B$.  The trajectory
contributes to tunneling only if the ends of the trajectory are on
different sides of $\op_C$.  Thus, the Monte Carlo version of
\eq\nr{flux} becomes
\be
  \Gamma V  =  P_C^\epsilon
	\, \fr12
	\left \< 
         \bigg| \fr{\Delta \op(\alpha)}{\Delta t}\bigg| \times
         \bd^\alpha
        \right\>\,,
\la{mcrate1}
\ee
where we have defined $\bd^\alpha = 
\delta_{\rm tunnel}^\alpha/N^\alpha_{\rm crossings}$,
and 
\be
  P_C^\epsilon = 
  \fr1{ \epsilon P_{\rm symm.}}\,
  \int_{\op_C-\fr\epsilon2}^{\op_C +\fr\epsilon2} P(\op)d\op\,.
\ee
Here we have substituted the $\delta(\op-\op_C)$-function in
\eq\nr{flux} with a small but finite width $\epsilon$; in practice, it
is much easier to generate configurations in this ensemble than with a
sharp $\delta$-function. 
The derivative $d\op/d t$ is again evaluated
at the initial configuration $\phi$, and the expectation value
$\<\cdot\>$ is over the configurations and trajectories.

The evolution backwards in time means that we invert the (possible)
initial momenta $\pi\rightarrow -\pi$ and evolve the system using the
standard equations of motion; the time is just interpreted as $-t$.
If the dynamics does not involve momenta (for example Langevin or
Glauber), then the evolution backwards is just another realization of
the standard forward evolution.  The detailed balance condition
\nr{balance} guarantees that the trajectory generated by
this backwards evolution can be re-interpreted as a representative
forward evolving trajectory.

The expression \nr{mcrate1} is straightforward to use in simulations.
However, we simplify it here a bit further: for the physical process
we are considering in this work, it turns out that $|
\Delta\op(\alpha)/\Delta t|$ is almost completely uncorrelated with
the ``global'' properties of the trajectory $\alpha$, that is, how
many crossings of $\op_C$ it makes, or whether it finally contributes
to tunneling.  This implies that to good approximation 
we can evaluate the expectation
values of the two terms inside the angle brackets separately:
\be
  \Gamma V  \simeq 
        P_C^\epsilon
        \fr12
	\left \< 
        \bigg| \fr{\Delta \op}{\Delta t}\bigg| \right\> \,
        \left \< 
        \bd
        \right\>\,.
\la{mcrate2}
\ee
Both of the expectation values above are thermal averages over all
trajectories which go through $\op_C$.  This is the form of the rate
equation given in Ref.~\cite{ewbubble}.  The advantage here is that
$\langle | \Delta \op / \Delta t | \rangle$
is a `local' quantity, and its evaluation does not
require the full knowledge of the trajectory.  Indeed, for the cubic
anisotropy model, with the order parameter and the real-time dynamics we
shall consider in this work, it is possible to evaluate it 
analytically, see Sec.~\ref{sec:dphidt}.
The decomposition of the expectation value in \nr{mcrate2} is a
natural consequence of the fact that the critical droplets are large
objects, and thus must involve 
a large number of microscopic variables.
The evolution of the UV-modes of these
variables is largely uncorrelated across the droplet.  The decomposition
holds strictly in the zero lattice spacing limit, where $|d\avphi/dt|$
is UV dominated.  However, if one
applies the method described here to tunneling processes where only
relatively few microscopic degrees of freedom participate, one cannot
rely on the decomposition and the equation \nr{mcrate1} must be used
instead.

\subsection{Independence on the order parameter}\la{sec:rateop}

In the discussion above we made extensive use of the critical droplet
order parameter value $\op_C$, where the free energy $F(\op)$ has a
local maximum.  However, we could have used any other value of $\op$
in the neighborhood of the free energy maximum --- we only have to
make sure that the whole of the bulk of the symmetric phase
probability remains on the symmetric phase side of it.  Intuitively
this is clear.  What Eq.~(\ref{flux}) actually measures is the number of
trajectories going from $\op_S$ to $\op_B$; and this must
be the same, no matter what point in between we decide to
intercept them.  For example, we may choose a new value $\op'$,
somewhat to the symmetric phase side of $\op_C$, to be our new
starting point.  The probability factors in the integrals \nr{flux} or
\nr{mcrate1} increase (the free energy is lower), but this is exactly
compensated for by the decrease in the average value of the projection
$\delta_{\rm tunnel}^\alpha$: most of the trajectories which intersect
$\op'$ will both start and end in the symmetric phase and hence do not
contribute to the tunneling.  However, if we want to get as accurate
a result as possible for a given amount of cpu-time, it pays to choose
the starting value as close to $\op_C$ as possible.

Very significantly, the rate calculated with \nr{flux}--\nr{mcrate2} is
{\em independent of the choice of the order parameter itself}.  This
is in contrast to the probability distribution $P(\op)$ alone (which
can be used to calculate the free energy of the critical droplet, see
Sec.~\ref{sec:droplet}).  The requirement the order parameter has
to fulfill is that it separates the symmetric and broken phases and
resolves the potential barrier between them to a sufficient degree, in
order for one to be able to choose the points corresponding to
$\op_S$, $\op_C$ and $\op_B$.  Provided that the hierarchy in free
energies $F(\op_C) \gg F(\op_S), F(\op_B)$ is satisfied, the order
parameters yield equal results, up to exponentially suppressed
corrections, of order $\exp[(F(\op_S)-F(\op_C))] \ll 1$.

However, not all order parameters are created equal: the best
parameter has as small fluctuations in the bulk phases as possible,
while separating the two phases as strongly as possible.  Large
fluctuations will reduce the resolving power of the order parameter: for
example, we may have a droplet which is slightly smaller than the true
critical droplet, but a fluctuation in the bulk may still bring the
order parameter up to $\op_C$.  These fluctuations will typically
increase the probability $P(\op_C)$, and again, this increase is
exactly cancelled by a decrease in $\delta_{\rm tunnel}^\alpha$: the
trajectories evaluated from these configurations are less likely to
lead to tunneling.  Thus, the better the order parameter, the
larger the fraction of trajectories which lead to tunneling.
Especially important is that the fluctuations of the order parameter
in the symmetric phase should be very small, since more than 85\% of
the lattice volume in a critical droplet configuration will remain in
the symmetric phase, outside the droplet.

\section{Cubic anisotropy model} \la{sec:cubic}

For concrete calculations in this work we use the cubic anisotropy
model in 3 dimensions.  The reasons for choosing this are the
following: (1) it has a radiatively generated first order phase
transition, the strength of which can be adjusted continuously; (2) it
is a (superficially) simple model and relatively easy to study on the
lattice; (3) it is a continuum field theory, and we can test whether
our nucleation calculation gives consistent results as the continuum
limit is taken on the lattice.  Furthermore, recently the validity of
Langer's theory of nucleation has been questioned in radiatively
generated transitions, using exactly this model as an example
\cite{StrumiaTetradis}.

\subsection{Definition of the model}

The cubic anisotropy model is a model with two scalar fields $\phi_1$
and $\phi_2$, with quartic interactions respecting the discrete
symmetries $\phi_1 \leftrightarrow \phi_2$ and $\phi_1 \leftrightarrow -
\phi_1$.
The thermodynamics of the model is defined by the partition function
(We use here ``particle physics'' units where [length] = [time] =
[energy]$^{-1}$ = [temperature]$^{-1}$),
\be
  Z = \int {\cal D} \phi_1 {\cal D} \phi_2 \, \exp -H_0[\phi] \,,
\la{partition}
\ee
where the 3-dimensional action is
\be
  H_0[\phi] = 
  \int d^3 x 
  \left[ \sum_{a=1,2} \left( \fr12 \sum_{i=1}^3 ( \partial_i\phi_a )^2 
+ \fr{m^2}{2} \phi_a^{2} 
+ \fr{\lambda_1}{24} \phi_a^4 \right)
+ \fr{\lambda_2}{4}  \phi_1^2 \phi_2^2  \right]
\,.
\la{energy}
\ee
The scalar fields $\phi_1, \phi_2$ have dimensions
of $\mbox{(length)}^{-1/2}$, and the coupling constants
$\lambda\sim\mbox{(length)}^{-1}$.  In the case that the 3 dimensional
theory arises from dimensional reduction of a 3+1 dimensional,
relativistic scalar field theory, the 3-D quantities are related to the
3+1 dimensional quantities through $H_0=H_{\rm 4D}/T$, $\phi = \phi_{\rm
4D} T^{1/2}$, $\lambda = \lambda_{\rm 4D} T$, and $m^2 = m^2_{\rm 4D} +
O(T^2)$.  Besides rescaling the 3 dimensional parameters, changing the
temperature of the 3+1 dimensional system adjusts the value of $m^2$,
which is why we often equate lowering $m^2$ with cooling.

The mass parameter $m^2$ drives the transition:
when $m^2$ has a large enough positive value, the system is in the
symmetric phase $\ev{\phi_1} = \ev{\phi_2} = 0$, whereas with large
and negative $m^2$ we are in the broken phase where $|\ev{\phi_1}| +
|\ev{\phi_2}| > 0$.  At the mean field level (``tree level'' in
particle physics parlance) the transition between the symmetric and
broken phases is of second order and happens at $m^2 = 0$.  Also at
the mean field level, the stability of the theory requires $\lambda_1
> 0$ and $\lambda_2 > - \lambda_1/3$.
We can immediately recognize some special values for the couplings
$(\lambda_1,\lambda_2)$: first, when $\lambda_2 = \lambda_1/3$, the 4th order
term in the energy functional becomes
$(\lambda_1/24) (\phi_1^2 + \phi_2^2)^2$.  Thus, the
theory has then an O(2) symmetry under rotations of the fields
$(\phi_1,\phi_2)$, and as we vary $m^2$, there is a second order O(2)
phase transition. Another special point is
$\lambda_2 = 0$, where we have two uncoupled $\phi^4$ models. Now the
transition is again of second order, of Ising model universality.
We also note that the redefinition of the fields 
\be
  (\phi_1 , \phi_2) \rightarrow 
  \fr{1}{\sqrt{2}} (\phi_1 + \phi_2, \phi_1 - \phi_2)
\ee
maps \eq\nr{energy} onto itself, with
\be
  m^2 \rightarrow m^2, ~~~ 
  \lambda_1 \rightarrow \fr12 (\lambda_1 + 3\lambda_2), ~~~
  \lambda_2 \rightarrow \fr12 (\lambda_1 - \lambda_2)\,.
\la{lambdamapping}
\ee
The O(2) invariant point $\lambda_2 = \lambda_1/3$ is a fixed point of
this transformation, and the region $\lambda_2 > \lambda_1/3$ is
mapped on $\lambda_2 < \lambda_1/3$.  This implies that the point $\lambda_1 =
\lambda_2$ is equivalent to $\lambda_2 = 0$, and it too corresponds to two
uncoupled $\phi^4$ theories with Ising universality.  

The behavior of the system at general values of the couplings has
been studied by renormalization group methods by Rudnick \cite{Rudnick} and
Arnold and Yaffe \cite{AYcubic}.  Their analysis indicates the following
behavior:
\begin{enumerate}\itemsep=0cm

\item
If $0 < \lambda_2 < \lambda_1$, the theory will flow to the O(2) fixed
point at large distances, and so will have a second order transition
with O(2) criticality.  In the broken phase there is a Goldstone mode,
corresponding to rotations of the $(\phi_1,\phi_2)$-vector.

\item
If $\lambda_2 = 0$ or $\lambda_2 = \lambda_1$, we have two uncoupled
$\phi^4$ theories, and the transition is of Ising type.

\item
If $\lambda_2 < 0$ or $\lambda_2 > \lambda_1$, the theory does not
flow to any weakly coupled infrared stable fixed point, and so might
be expected to have a first order phase transition.  This is indeed
the case, as shown by perturbative analysis \cite{Rudnick,AYcubic} and
lattice simulations \cite{ASYZcubic}. 

\end{enumerate}
The first order transition in the last case is generated by radiative
effects; in order to observe it perturbatively, one must consider the
first loop correction (Coleman-Weinberg effect
\cite{ColemanWeinberg}).  Using the $\epsilon$ expansion around
$d=4-\epsilon$ dimensions, Arnold and Yaffe
\cite{AYcubic} calculated the effective potential up to
next-to-next-to-leading order in $\epsilon$.  In the limit $\lambda_2
\gg \lambda_1$, the transition becomes strong and the perturbative
analysis should be accurate.  There are 4 degenerate broken phases in
the region of the first order phase transition.  If we choose
$\lambda_2 > \lambda_1$, in the broken phase the $(\phi_1,\phi_2)$
-doublet acquires an expectation value along one of the principal
axes: $(\pm v,0)$ or $(0,\pm v)$.  

In this work we are interested in strong first order transitions.
Thus, we choose $\lambda_2 = 8\lambda_1$, which is seen to give a very
strong transition.

\subsection{Cubic anisotropy model on the lattice}\la{cubiclat}

We discretize the theory on the lattice using an ${\mathcal O}(a^2)$
accurate lattice action, where $a$ is the lattice spacing.  The
action can be written as
\ba
H_0
  & = & \sum_x \Bigg[ 
	\frac{Z_\phi}{2} \left( - \phi_1 \nabla^2_{\rm L} 
	\phi_1 - \phi_2 \nabla^2_{\rm L} \phi_2 \right) 
	+ \frac{Z_m (m^2 + \delta m^2)}{2}  
	\left( \phi_1^2 + \phi_2^2 \right) + \nonumber \\ & &
	\qquad + \frac{(\lambda_1 + \delta \lambda_1)}{24} 
	\left( \phi_1^4 + \phi_2^4 \right)
	+ \frac{(\lambda_2 + \delta \lambda_2)}{4} 
	\phi_1^2 \phi_2^2  \Bigg] \, ,
\la{latticelagrangian}
\ea
where we have rescaled the lattice fields and couplings to be
dimensionless: $\phi_{\rm latt.} = \phi_{\rm cont.} a^{1/2}$,
$\lambda_{\rm latt.} = \lambda_{\rm cont.} a$, and $m^2_{\rm latt.} =
m^2_{\rm cont} a^2$.  $\delta$-terms and $Z$-factors are additive
and multiplicative renormalization constants, respectively.  They are
necessary to match the behavior of the lattice theory to the continuum
theory without $O(a^2)$ errors, and their calculation is detailed in the
appendix.  Eliminating such errors also requires a lattice Laplacian
which includes next-neighbor couplings, 
\be
\nabla^2_{\rm L} \phi(x) = - \frac{15}{2} \phi(x) + \frac{4}{3} \sum_i
	\Big( \phi(x+i) + \phi(x-i) \Big) - \frac{1}{12} \sum_i \Big( 
	\phi(x+2i) + \phi(x-2i) \Big) \, . \la{imp_Laplace}
\ee
Furthermore, the operator insertion 
$\< \phi^2 \> = \< \phi_1^2 + \phi_2^2 \>$, which we will need in what
follows, is additively and multiplicatively corrected,
\be
a  \< \phi^2 \>_{\rm contin} = Z_{m}\< \phi^2 \>_{\rm latt} 
	- \delta \< \phi^2 \> \, .
\la{phi2lattcont}
\ee

\subsection{Real time dynamics}\la{sec:dyn}

To define the nucleation {\em rate\,}, we must specify
a dynamical prescription for the time evolution.  The rate will depend
on the choice, which, after all, determines what physical system we are
describing.  Any prescription which preserves the canonical distribution
is permitted.  The most common
evolution prescriptions in field theories are the Hamiltonian and the
Langevin dynamics.  Both are perfectly valid evolution dynamics for
classical thermal field theories; for our case the Hamiltonian
dynamics has the further conceptual advantage that it also describes
the evolution of a {\em quantum} theory at high temperatures to
leading order in coupling constants \cite{bodek_h,AartsSmit}.

In this work we shall consider a one-parameter class of evolution
dynamics, the {\em Hamiltonian stochastic equations\,}
\cite{Horowitz,RyangSaitoShigemoto}
\ba
  \partial_t \phi_a(x,t)  &=&   \pi_a(x,t)  \nonumber \\
  \partial_t \pi_a(x,t)   &=&  - \fr{\delta H_0}{\delta \phi_a} + 
	\gamma \pi_a(x,t) + \xi_a(x,t)\,. \la{eqm}
\ea
Here $H_0$ is given by \eq\nr{energy}, $\pi_a$ are the momentum
variables of fields $\phi_a$, $\gamma \ge 0$ is a ``friction
coefficient,'' and $\xi_a(x,t)$ is Gaussian noise, which satisfies the
stochastic condition
\be
   \< \xi_a(x,t)\xi_b(x',t') \> = 
  2\gamma \delta_{ab} \delta^3(x-x')\delta(t-t')\,.
\ee
The equations of motion \nr{eqm} thermalize the system (if $\gamma
> 0$), that is, at large times the probability distribution approaches
the Gibbs distribution $p\propto \exp(-H[\phi,\pi])$, where
\be
  H[\phi,\pi] = \int d^3x \fr12 (\pi_1^2 + \pi_2^2) + S_0[\phi] \,.
\ee
Thus, the evolution also preserves the correct thermodynamics of the
``static'' theory, \eq\nr{partition}.  

If we choose $\gamma=0$ in \eqs\nr{eqm} the evolution becomes the
standard canonical Hamiltonian evolution, with canonical momenta
$\pi_a$ and conserved energy $H$.  On the other hand, if $\gamma$ is
very large, the equations of motion can be reduced to the familiar
fully diffusive Langevin equation for $\phi$, simply by eliminating
$\pi$ and neglecting the term (small in this limit) $\partial_t^2
\phi$.  Thus, by adjusting $\gamma$ we can continuously
adjust the coupling of the system to an external ``heat bath.''

For the case of droplet nucleation at finite volume the presence of
noise improves the finite volume and lattice spacing scaling of the
dynamics.  This is because the transition has
a large latent heat.  A growing/shrinking droplet will release/absorb a
significant amount of energy.  Under microcanonical Hamiltonian
dynamics the temperature in a finite system will correspondingly
increase/decrease.  In an infinite volume, the released energy would
be transported rapidly away from the proximity of the droplet by
hydrodynamical flows\footnote{Heat flow behaves very differently in a
plasma without nonzero conserved quantities than in more familiar
condensed matter or convective settings, see for instance
\pcite{Jeon}.}, and the temperature would remain close to
constant.  As the lattice spacing becomes smaller, the heat capacity of
the system grows, and the latent heat again becomes less important.
But at finite lattice spacing and volume, the latent heat remains
indefinitely under Hamiltonian dynamics, modifying the time development.
A finite $\gamma$ in \eqs\nr{eqm} absorbs the latent heat, allowing the
system to more rapidly approach the infinite volume behavior.  It does
this efficiently for $\gamma \gsim 1/L$, in which case the
momenta thermalize completely in time $\tau \lsim L$.  In 
Sec.~\ref{sec:rate} we check the dependence of the nucleation
rate on $\gamma$, and we find that we can vary $\gamma$ over
a large range without significantly affecting the results.
This means that, in practice, the nucleation rate has a fairly weak
dependence on the specifics of the real time dynamics.

\subsection{Real time evolution on the lattice}

The discretization of the equations of motion \nr{eqm} requires some
care, because certain errors lead to time evolution which respects
a slightly different ensemble of configurations than the Gibbs one.
This is important, because slight thermodynamic changes dramatically 
affect the critical bubble behavior. 
The easiest safe update at nonzero $\gamma$ is to apply Hamiltonian
evolution, stopping every $\Delta t \ll a$ to apply a partial momentum
refresh, 
\be
     \pi_a(x,t+0) = (1 - \varepsilon^2)^{1/2}\,
     \pi_a(x,t-0) + \varepsilon \eta_a(x,t)\, ,
\la{vareps}
\ee
with $\epsilon^2 = \left( 1 - \exp(-2\gamma \Delta t)\right)$.  This
refresh exactly preserves the Gibbs distribution.  We perform $\Delta t$
of Hamiltonian evolution by a single step of the fourth order
Runge-Kutta algorithm \cite{numrec}, which has errors suppressed by
$(\Delta t/a)^5$.  This procedure 
reproduces Eq.~(\ref{eqm}) with errors
of order $O([\Delta t/a]^2)$, $O(\epsilon^2)$,
but what is more important, it maintains the Gibbs distribution to
$O([\Delta t/a]^4)$ errors.  We use $\Delta t/a = 0.05$, which proves
more than adequate, unless we want to study extremely large $\gamma$.

\section{Numerics and thermodynamic results}\la{sec:numerics}

\subsection{Order parameter}\la{sec:orderp}

The cubic anisotropy model has a true order parameter, which
measures the breaking of the global symmetry:
\be
\avphione = \sqrt{(\overline \phi_1)^2 + (\overline \phi_2)^2} \, , \qquad
\overline\phi_1 = \frac{1}{N^3} \sum_x \phi_1 \, .
\la{trueorderp}
\ee
Here $N^3$ is the volume in lattice units.
However, a better observable for our purposes is
\be
  \avphi = \fr1{N^3} \sum_x (\phi^2_1 + \phi^2_2) \, .
\la{avphi}
\ee
This is clearly not an order parameter in the sense that it
would be zero in one phase and non-zero in another.  However, it has
different expectation values in the symmetric and broken phases, which
is sufficient for our purposes.  The reasons why we prefer $\avphi$ over
$\avphione$ are that (a) it is much more economical to use, and (b) it
resolves the critical droplet more accurately.

Let us first consider the cost.  Note that $\avphi$ is multiplied
by $m^2$ in the Hamiltonian \nr{energy}.  Assuming that we have
measured the order parameter distribution at $m^2 = m_1^2$,
\be
  P_{m_1^2}(\avphi) \propto 
	\int d\phi' \exp[-H_{m_1^2}(\phi')]\,
	\delta(\phi'^{\,2}_{\rm av}-\avphi)\,,
\ee
we obtain the distribution at any other $m_2^2$ simply by {\em
reweighting:}
\be
  P_{m_2^2}(\avphi) \propto \exp[(m_1^2 - m_2^2)\avphi] P_{m_1^2}(\avphi)\,.
\la{reweight}
\ee
The reweighting can be done without any significant loss of accuracy:
if we have determined $P_{m_1^2}$ to a good accuracy in some
$\avphi$-range of interest, reweighting gives $P_{m_2^2}$ with the
same relative error $\delta P(\avphi)/P(\avphi)$ in this
range.\footnote{We can also reweight $P(\avphione)$ with respect to
$m^2$ (or any other order parameter distribution, for that matter);
however, in this case the errors increase dramatically when $m^2$ is
outside a narrow range around the value where the original measurement
was done.}
Since the calculation of $P(\avphi)$ comprises the major part of
the total computational effort, it is quite significant that 
we can perform the measurement once, and use it for all $m^2$ values.  
Of course, separate calculations are still required for different lattice
volumes and lattice spacings.

As discussed in Sec.~\nr{sec:rateop}, in order to be able to
resolve the droplet as well as possible, the fluctuations of the order
parameter in the {\em symmetric phase}%
\footnote{%
	In general the argument applies to the phase the system is
	tunneling out of.  In particular, $\avphi$ would be a very
	bad order parameter if we were studying tunneling from
	the superheated, broken phase into the symmetric one; we would 
	want to choose instead some measurable with small fluctuations
	in the broken phase.
	} %
should be as small as possible
--- this is because more than 85\% of the volume of the lattice is in
the symmetric phase in critical droplet configurations.  Here $\avphi$
is a clear winner over $\avphione$.  This can be seen by comparing the
(width)$^2$ of the symmetric phase distributions, i.e. the
susceptibilities:
\ba
  \chi(\avphione) 
       &=& \fr1V \int d^3x \< \avphione(x)\avphione(0) \> 
	= \fr{2}{V m^2} \\
  \chi(\avphi) &=& 
	\fr1V \int d^3x \< \avphi(x) 
	\avphi(0)\>_{\rm connected} \nonumber \\
       &=& \fr4V \int \fr{d^3p}{(2\pi)^3} \fr{1}{(p^2+m^2)^2} 
	= \fr{1}{2\pi V m}\,.
\ea
Here $m^2$ is the (mass)$^2$ of $\phi$ in the symmetric phase.  Since
the tree-level transition happens at $m^2=0$, the value of $m^2$ at the
transition must be small, of order $\lambda^2$.  The relevant
quantity at the transition is the width of
the symmetric phase distribution compared against
the broken phase expectation value; 
using 1-loop effective potential, see Eq.~(\ref{v1}), we obtain
the relative ``figure of merit''
\be
  \fr{\chi^{1/2}(\avphione)/v}{\chi^{1/2}(\avphi)/v^2} \approx 13
	{\rm \; for \;} \lambda_2 = 8 \lambda_1 \, , 
\ee
where $v = \<\avphione\>$ in the broken phase.  Thus, the symmetric
phase fluctuations of $\avphione$ are more than 10 times larger than
$\avphi$, when normalized to the broken phase expectation values.
Naturally, the final word comes from the lattice simulations, where
$\avphione$ is indeed observed to give a $\sim 10$ times wider symmetric
phase peak than $\avphi$.  The width of the symmetric peak of the the
$\avphione$-distribution is actually larger than the critical droplet
value $\phi_{{\rm av},C}$ for the parameters we are using in this
work.  Thus, $\phi_{{\rm av},C}$ does not have the necessary
resolution power, while we can use $\avphi$ without problems.  A
higher power of $\phi$, for example, $\phi^4_{\rm av}$, might give
an even narrower symmetric phase peak on a coarse lattice spacing, but
in the continuum it
would have a UV divergent susceptibility, unlike $\avphi$, so taking
the continuum limit might prove problematic.  Such order parameters
also fail the economy criterion.

\subsection{Multicanonical method}\la{sec:multi}

In order to be able to calculate the droplet nucleation rate, our task
now is to determine $P(\avphi)$ in the droplet regime,
$\vfrac < 0.15$.  Since the probability density can
vary by $\sim \exp(100)$ or more in this range, normal Monte Carlo
simulations, where the configurations are sampled with probability
$p\propto\exp-H$, are completely out of the question.  The {\em
multicanonical} method \cite{BergNeuhaus} was originally developed
for just this kind of problems.  In multicanonical sampling
the configurations are chosen with a modified weight
\be
  p_{\rm muca} \propto \exp[-H_{m^2} + W(\avphi)]\,.
\ee
$W(\avphi)$ is a {\em weight function}, which is carefully chosen so
that the probability distribution of $\avphi$ measured from the
multicanonical Monte Carlo simulation, $P_{\rm muca}(\avphi)$, is
approximately constant in the range of interest.  The canonical
distribution is then obtained by reweighting with the weight function:
\be
  P_{m^2}(\avphi) \propto e^{-W(\avphi)} P_{\rm muca}(\avphi).
\ee
This can be further reweighted to other values of $m^2$ using
\eq\nr{reweight}.  

An optimal choice for $W$ is $-\log P_{m^2}(\avphi)$, which is just
the quantity we want to calculate with the multicanonical simulation!
The weight function need not be exactly the ideal one, but it must not
deviate from it too much or the multicanonical simulation becomes very
inefficient.  This chicken-and-egg problem can be resolved by first
calculating $W$ with an automatic iterative feedback procedure
\cite{mssmsim,Berg2}.  We use
a variation of the one presented in Ref.~\cite{mssmsim}.  This
method yields progressively better approximations for $W$, until
sufficient accuracy is reached.  The final $W$ can then be used in a
multicanonical simulation, which finally gives us $P(\avphi)$.
A detailed description of the application of the multicanonical method
to the problem of the bubble nucleation in the electroweak theory can
be found in \cite{ewbubble}.

In the multicanonical phase of the simulations we are at liberty to
choose as efficient an update algorithm as possible. (In the real-time
runs the evolution equations fix the update.)  We use a
mixture of site by site heat bath and a non-local multi-grid
over-relaxation.

\subsection{Results: thermodynamics}\la{sec:thermo}

Before tackling the droplet nucleation rate calculation, we want
to study basic thermodynamic quantities associated with the
phase transition: the transition ``temperature'' $m^2 = m_c^2$, the
``latent heat'' $\lat$, and the tension
of the interface between the symmetric and broken phases $\sigma$.
We want the above quantities in the true thermodynamic limit,
that is, we extrapolate to infinite volume and to zero lattice spacing.
All of the simulations have been done with $\lambda_1/\lambda_2 = 1/8$,
which guarantees a strong first order transition.

In this part of the analysis we use up to 6 different lattice spacings
$a\lambda_2 = 1.5$, 2, 2.5, 3, 3.5 and 4 (all dimensionful quantities
are given in terms of $\lambda_2$).  For every lattice spacing
we perform simulations using a series of lattice sizes; 
all in all, we have 64 simulations at different lattice spacings and
volumes.  Since the transition is strong, all of the simulations
described here are multicanonical (see subsec.~\ref{sec:multi}), with the
weight function optimized for the whole order parameter range from the
symmetric to the broken phase.

\begin{figure}[tb]
\centerline{
\epsfxsize=9cm\epsfbox{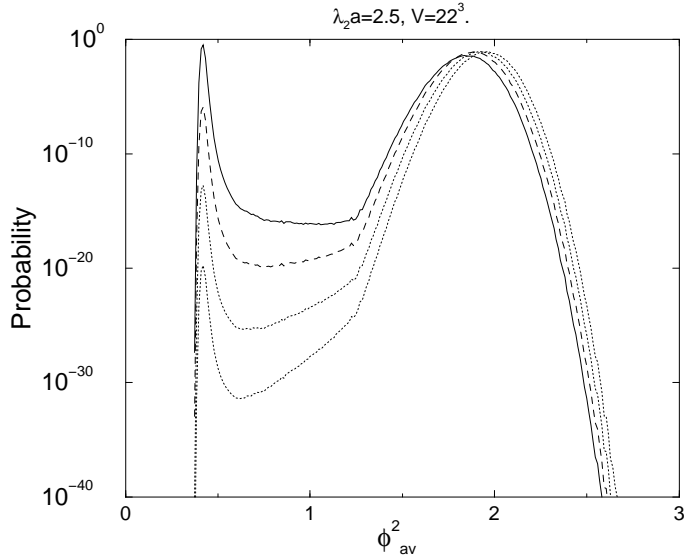}
}
\caption[a]{Reweighting of the probability distribution $P(\avphi)$
to different values of $m^2$.  The transition value $m_c^2$ is the 
one where the symmetric and broken phase peaks have equal volume.}
\label{probdist}
\end{figure}

\paragraph{The transition point $m_{c}^{2}$:}

\begin{figure}[tb]
\centerline{
\epsfxsize=7cm\epsfbox{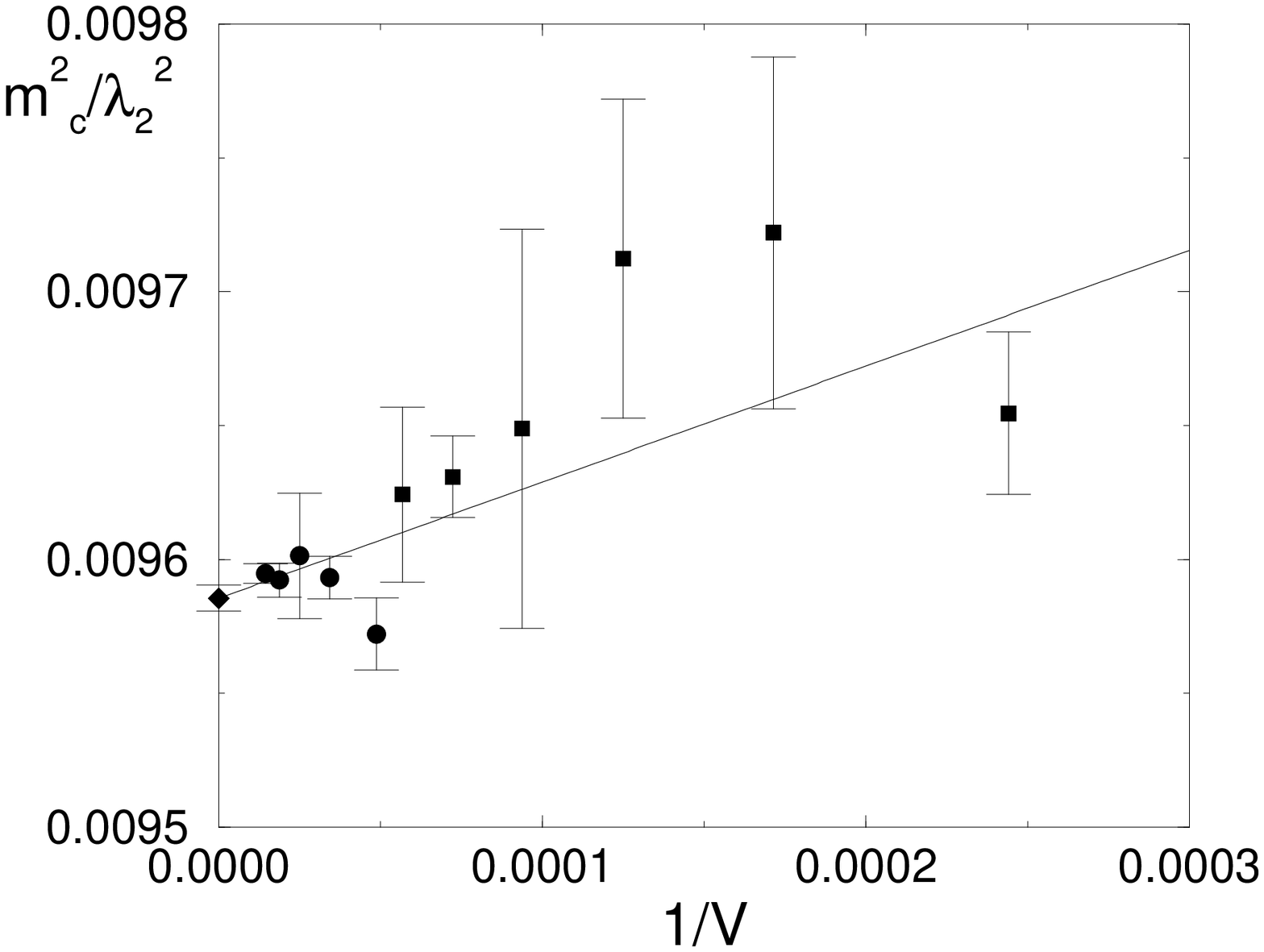}
\epsfxsize=7cm\epsfbox{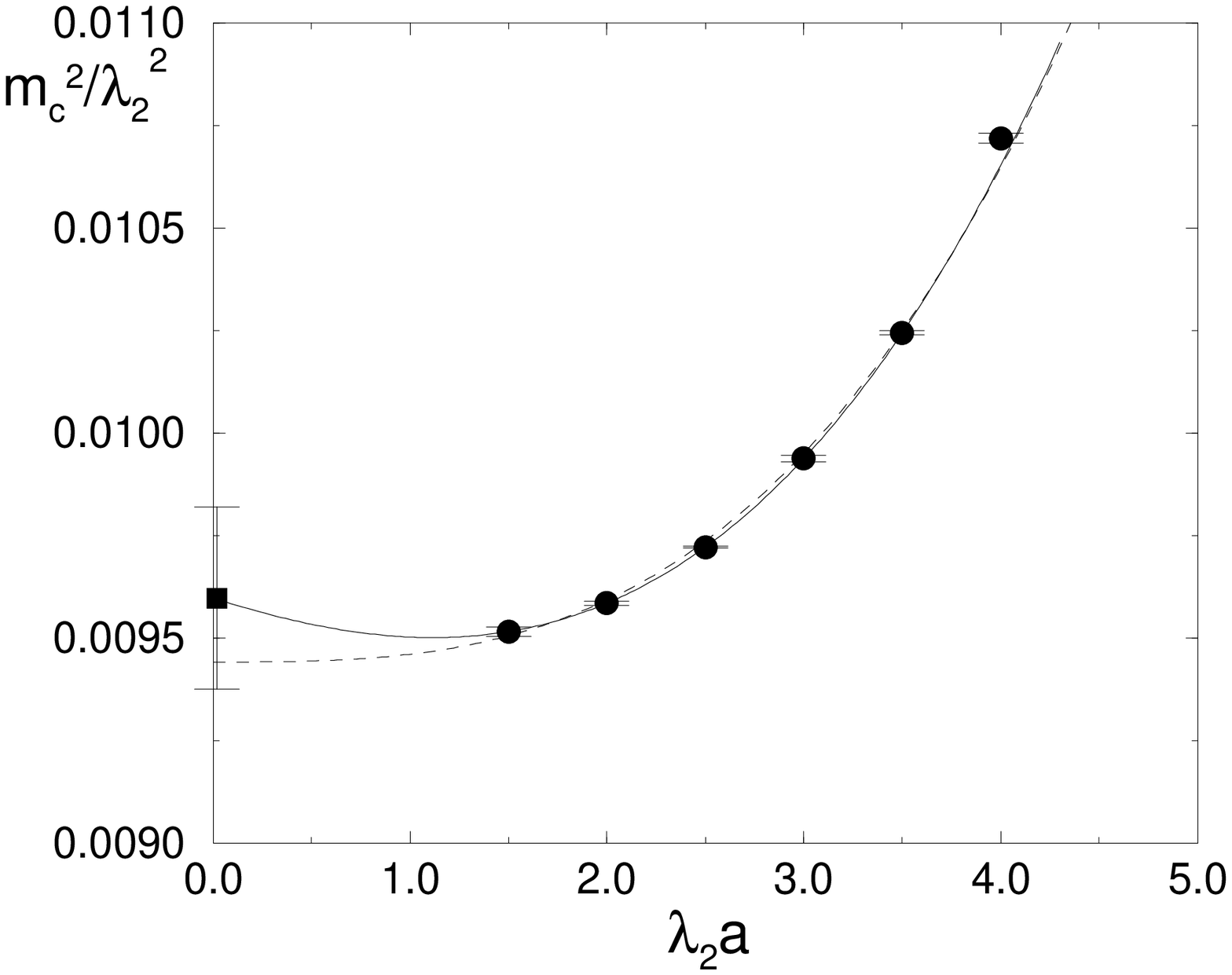}
}
\caption[a]{
Transition value of $m^{2}$ extrapolated to infinite volume 
at $\lambda_{2}a=2$ (left), and to zero lattice spacing (right).
The solid line shows a 3rd order polynomial fit,
and the dotted line is a fit of form $c_0 + c_3(\lambda_2 a)^3$.
}
\label{betaa}
\end{figure}

We define the transition value of $m^2$ to be the value where the two
peaks of the probability distribution $P(\avphi)$ have {\em equal
weight}, i.e. the value where the volumes of the symmetric and broken
phase peaks are equal (see \fig\ref{probdist}).  More precisely, the
symmetric and broken phase probabilities are the integrals of the
distribution to the right and to the left of a separating value
between the peaks; since the probability is exponentially suppressed
here in sufficiently large volumes, the precise choice of the
separation value is exponentially unimportant.
For each lattice spacing $a$, we determine the transition point $m_c^2$
using a series of volumes and extrapolate the results to infinite
volume.  The extrapolation is linear in $1/V$, as long as the volumes
have similar geometry; an example of this at $a = 3/\lambda_2$ is
shown on the left panel of \fig\ref{betaa}.  

The infinite volume points are in turn extrapolated to the continuum
limit $a\rightarrow 0$.  This is shown on the right panel of
\fig\ref{betaa}.  
Since we know the additive mass counterterms only to order $O(a^0)$
(see appendix~\ref{sec:app}), we extrapolate to the continuum limit
using an ansatz $c_0 + c_1 x + c_2 x^2 + c_3 x^3$, where $c_i$ are
fitted constants and $x=a\lambda_2$.  The final result is
\be
    m^2_c(\mu=\lambda_2)/\lambda_2^2 = 0.0096 \pm 0.00022
\ee
We note that the supercooling $\delta m^2$ can be determined much more
accurately than $m_c^2$: in differences of $m^2$ 
the additive counterterms cancel,
and the leading errors behave as $O(a^3)$.  At no stage of the nucleation
analysis do we need to know the absolute value of $m_c^2$.

In the fit the coefficients $c_1$ and $c_2$ are actually consistent with
zero, suggesting that the unknown counterterms are small.  We also show
a fit with these coefficients set to zero.

\begin{figure}[tb]
\centerline{
\epsfxsize=7cm\epsfbox{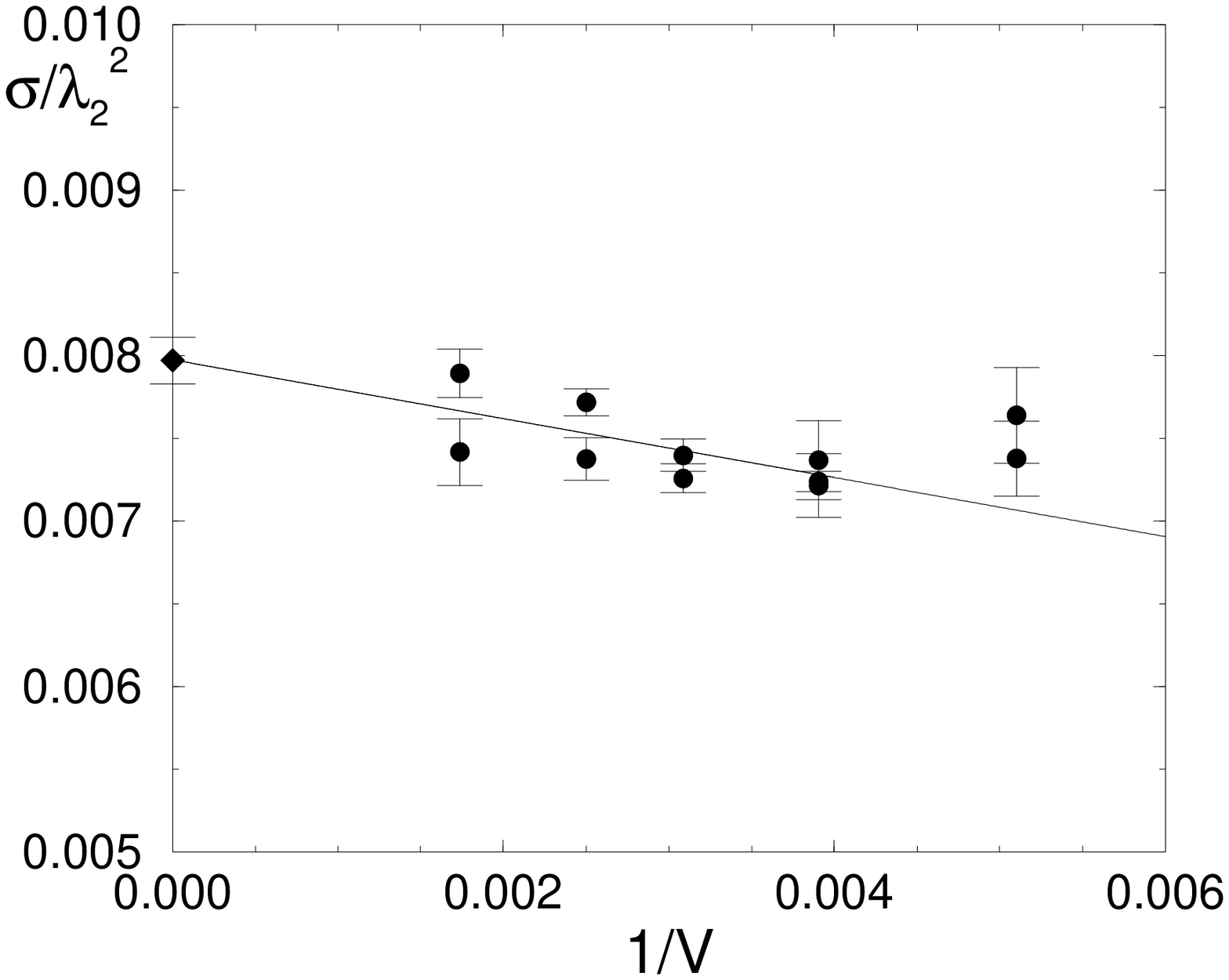}
\epsfxsize=7cm\epsfbox{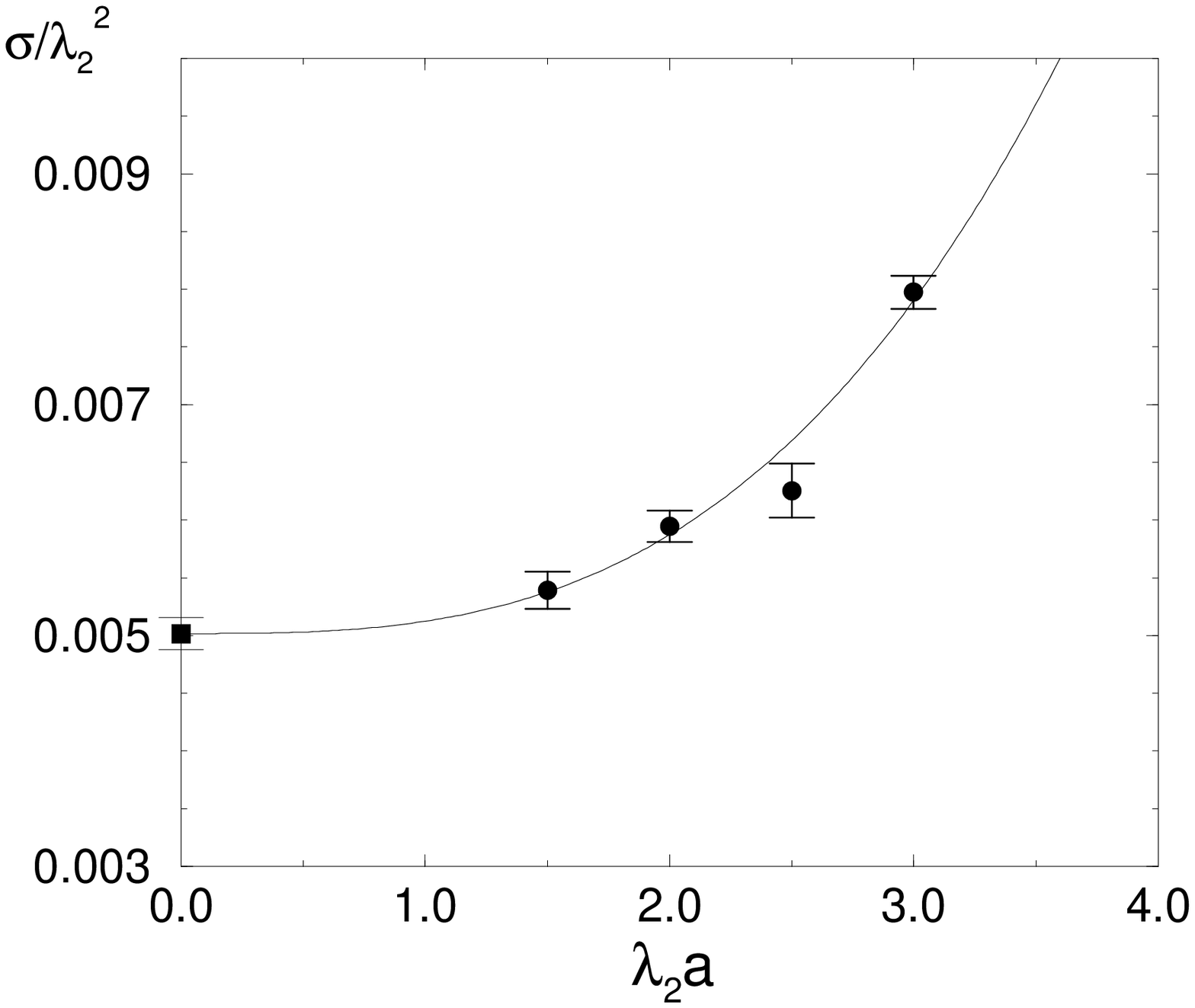} 
}
\caption[a]{
Surface tension $\sigma/\lambda^2_2$ extrapolated to 
infinite volume at $\lambda_{2}a=3$ (left), and to zero lattice
spacing (right).}
\label{tensa}
\end{figure}

\paragraph{The surface tension $\sigma$:}

The surface tension is the free energy per unit area of interface.  As
discussed in Sec.~\ref{sec:droplet}, when $m^2 = m_c^2$ the suppression of the
probability in the mixed phase (in large enough volumes) is caused
by the existence of phase interfaces.  Approximately half-way between
the symmetric and broken phases, the mixed phase configurations
consist of slabs of symmetric and broken phases, separated by two
approximately planar interfaces.  Thus,
the interface tension $\sigma$ can be obtained from \cite{Binder}
\be
  \sigma = 
	-\lim_{V\rightarrow\infty} \, 
	\fr{1}{2 L_x L_y} \log\fr{P_{\rm slab}}{P_{\rm peak}}\,.
\la{binder}
\ee
Here we have assumed that the interfaces are oriented parallel to
$(x,y)$-plane.  As for other quantities we have scaled $T$ out of
$\sigma$; a more conventional definition is
$\sigma_c = \sigma T$, so that $\sigma_c\times\mbox{area}$ is the free
energy of the interface (rather than the scaled free energy we use in
this paper).  At the ``equal weight'' point the symmetric
and broken phase heights are not usually equal; in this case $P_{\rm
peak}$ is the average of the two peaks.  However, here we determine
$\sigma$ from distributions which have been reweighted to ``equal
height'' $m^2$; in our case, the two definitions give completely
consistent results.

The extrapolation to infinite volume in \eq\nr{binder} can be
substantially improved by using finite volume corrections
\cite{Bunk,Iwasaki}:
\be
  \sigma a^{2} 
   =\frac{1}{2n_{x}n_{y}} \log{\Big(\frac{ P_{\rm peak}}{P_{\rm slab}}\Big)}
    + \frac{1}{n_{x}n_{y}}\Big(\frac{4}{3}\log{n_{z}} - \frac{1}{2}\log{n_{x}} 
    + \frac{1}{2} G + \tx{const.}\Big)\,.
\la{sigmaV}
\ee 
Here $L_i = n_i a$ and $\sigma_L = \sigma a^2$ is the tension in
lattice units.  The function $G$ interpolates between lattice
geometries.  For cubical volumes $G=\log 3$,
while for long cylinders $G=0$.  In our analysis we use strongly
cylindrical boxes, $L_z \gg L_x = L_y$.  This guarantees that the
interfaces form along the $(x,y)$ plane, and it also separates the two
interfaces much farther than a cube of the same volume.  On the left
panel of \fig\ref{tensa} we show the extrapolation of
$\sigma/\lambda^2_2$ (corrected with \eq\nr{sigmaV}) to infinite
volume at lattice spacing $\lambda_2 a = 3$.  The continuum limit
extrapolation is shown on the right panel.  Because of our lattice
improvement the leading errors are $O(a^3)$.  We obtain a good fit
using a third order ansatz $\sigma(a) = \sigma + c_3 a^3$,
with the result
\be
  \sigma/\lambda_2^2 = 0.0050\pm 0.0002\,.
\ee

\paragraph{The latent heat $\lat$:}

Since $m^2$ is our temperature parameter, we define
the latent heat using
\be
  \lat = \Delta\left(\fr{d f}{d m^2}\right)_{m^2 = m_c^2}
       = \fr12 (\< \avphi \>_{\rm broken} - 
                \< \avphi \>_{\rm symm.}) \,.
\ee
The difference between the symmetric and broken phase expectation
values is readily measurable from the distributions like the one in
\fig\ref{probdist}.  Our extrapolation procedure is the same as for
$\sigma$: 
\be
  \lat/\lambda_2  = 0.243 \pm 0.004\,.
\ee

\section{Results: the nucleation rate}\la{sec:rate}

We evaluate the rate of the droplet nucleation using \eq\nr{mcrate2}.
To recap the discussion in Sec.~\ref{sec:howto}, the calculation
consists of two separate stages:

\noindent
(1) the measurement of the probability
distribution of the order parameter, $P(\avphi)$, using multicanonical
simulations, and 

\noindent 
(2) the real-time evolution of the critical droplets, which gives
us $\dpdt$ and $\<\bd\>$.

\begin{table}[thb]
\centerline{\begin{tabular}{|r|l|l|}
\hline
 $a\lambda_2$  &  size/$a^3$ &  evolution at $\delta m^2/\lambda_2^2$ 
\\
\hline
    4.0        &  $60^3$ &                      0.001156\\
    3.5        &  $70^3$ &   \\
    3.0        &  $60^3$, $70^3$, $80^3$     &  0.0007--0.00125\\
    2.5        & $100^3$ &                      0.00094\\
    2.0        & $120^3$ &   \\
\hline
\end{tabular}}
\caption[a]{The lattice spacing and size where the droplet nucleation
rate has been calculated.  For each lattice the probability
distribution $P(\avphi)$ has been calculated once with a multicanonical
simulation.  The real-time evolution of the critical droplets are
calculated using the supercoolings shown on the third column.}
\la{tab:lats}
\end{table}

The lattice spacings and sizes used are shown in Table \ref{tab:lats}.
For each of the lattices we calculate $P(\avphi)$ {\em once}; as
explained in Sec.~\nr{sec:orderp} and
\nr{sec:multi}, the original distribution can be reweighted to
different values of supercooling $\delta m^2$.  On the other hand, the
real-time trajectories have to be calculated separately for each
supercooling.

\subsection{The probability of the critical droplets}\la{sec:probdist}

In order to fit the droplets comfortably inside the lattice, the
lattice volumes have to be substantially larger than the volumes used
in Sec.~\nr{sec:thermo}.  However, as discussed in Sec.~\nr{sec:droplet},
in this case we have to calculate the probability distribution
$P(\avphi)$ only in the ``droplet branch'' of the distribution, 
$0\le[\avphi - \avphiX{{\rm s}}] < 0.15\times[\avphiX{{\rm
b}}-\avphiX{{\rm s}}]$.  This guarantees that the distance of
the droplet from its periodic copies is large and we are a safe
distance away from the ``cylinder'' branch of the distribution. 
Furthermore, the restricted range of $\avphi$ reduces the
computational requirements in multicanonical simulations
dramatically, when compared with the full
range calculations in Sec.~\nr{sec:thermo}; using random walk
arguments, the reduction factor is $0.15^2 \sim 0.02$; in practice,
the reduction is even stronger than this.

\begin{figure}[tb]
\centerline{
\epsfxsize=8cm\epsfbox{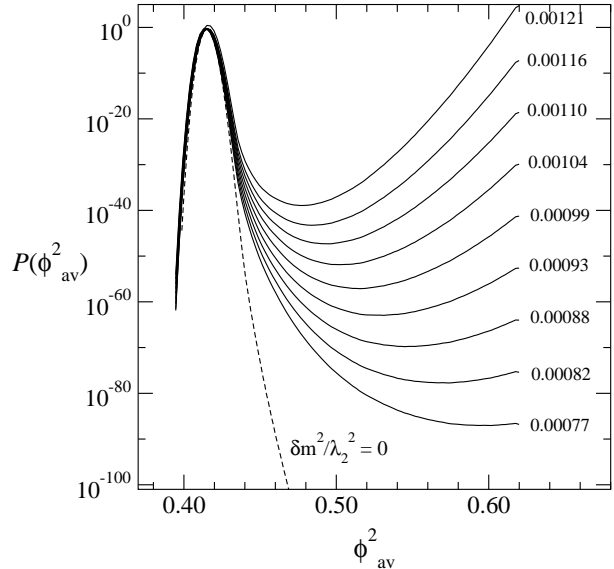}
}
\caption{The probability distribution $P(\avphi)$ at various 
values of supercooling $\delta m^2$, measured from $a\lambda_2 = 3$,
$80^3$ lattice.  
Here $\avphi$ is in lattice units, \eq\nr{phi2lattcont}.
The critical droplet value $\avphiC$ at each $\delta m^2$ 
is defined to be the location of the local minimum of $P(\avphi)$.
}
\label{rewprob}
\end{figure}

In \fig\ref{rewprob} we show the actual order parameter distributions
$P(\avphi)$, measured from the $a\lambda_2 = 3$, $80^3$ lattice, and
reweighted to supercooling values $\delta m^2/\lambda_2 =
0.00077$--0.00122.  The larger the supercooling, the smaller the
critical droplet, and less suppressed the droplet probability is.  
We always use {\em lattice\,} units for $\avphi$; since this is 
an ``internal'' quantity in the expression for the rate \nr{mcrate2},
there is no need to convert it to continuum units.

\begin{figure}[tb]
\centerline{
\epsfxsize=8cm\epsfbox{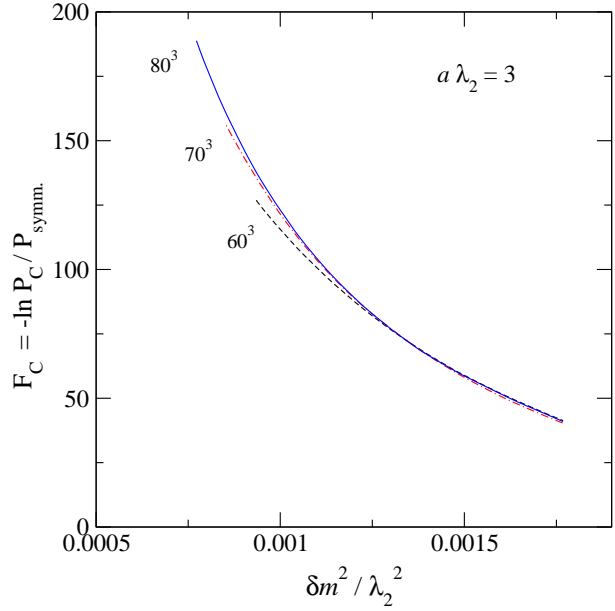}
}
\caption{
The critical droplet free energy $F_C$, as a function of
supercooling $\delta m^2$, measured from $\lambda_2 a = 3$ lattices of
size $60^3$--$80^3$.  
}
\label{fig:finitesize}
\end{figure}

We calculate the critical droplet free energy through $F_C\approx
-\ln(P(\avphiC)/P(\avphiX{{\rm s}}))$, where $P(\avphiX{{\rm s}})$
is the height of the symmetric phase peak of $P(\avphi)$.  Note that
this ratio is not $P_C^\epsilon$, which appears in the rate equation
\nr{mcrate2}; we chose to use this ratio because it is dimensionless,
whereas the probability factor in \nr{mcrate2} has dimensions of
$[\avphi]^{-1}$.  In \fig\ref{fig:finitesize} we show $F_C$ using
$\lambda_2 a = 3$ lattices of volumes $60^3$--$80^3$.  For large
values of supercooling $\delta m^2$, where the critical droplet is
small, we do not observe any significant finite volume effects.
However, when $\delta m^2$ is smaller, the critical droplet is large
and, in small lattices, the droplet feels the proximity of its
periodic copies.  For our smallest ($60^3$) lattice, this decreases
the droplet free energy significantly already at modest droplet sizes.
The curves shown in
\fig\ref{fig:finitesize} correspond to droplets which are well within
the maximum size determined by the 15\% rule discussed in
Sec.~\nr{sec:droplet}.  This finite size effect is caused by the finite thickness
of the droplet wall, and it should vanish exponentially as the lattice
size is increased.  Indeed, at large lattices we need to go to very
large droplets in order to observe any significant finite size
effects.

\begin{figure}[tb]
\centerline{
\epsfxsize=8cm\epsfbox{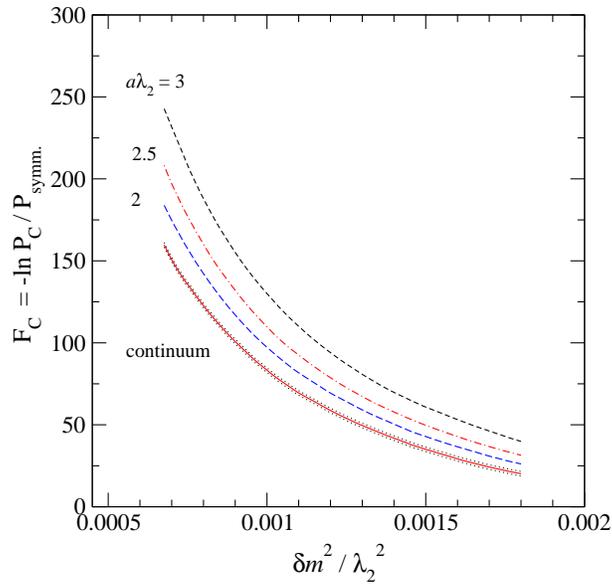}
}
\caption{
The critical droplet free energies measured from 
$\lambda_2 a = 3$, volume $80^3$; $\lambda_2 a = 2.5$, volume $100^3$, 
and $\lambda_2 a = 2$, volume $120^3$
lattices.  The continuum ($a=0$) curve is shown with an error
band.  
}
\label{fig:finitea}
\end{figure}

On the other hand, the lattice spacing effects are substantial for the
values of $a\lambda_2$ used: in \fig\ref{fig:finitea} we compare
$F_C$ from lattices of similar physical size, but $a\lambda_2 = 2$,
2.5 and 3.  
The larger lattice spacings $a\lambda_2 = 3.5$ and 4 have much larger 
finite $a$ effects and we discard these in our analysis.
For a fixed value of supercooling $\delta m^2$, $F_C$
decreases as the lattice spacing becomes smaller. 
The leading order errors are expected to be $\mathcal{O}(a^3)$; indeed, a good
fit to the 3 curves is obtained with a 2 parameter fit $c_0 + c_3 (a\lambda_2)^3$,
done independently for each $\delta m^2$.
The resulting continuum curve is shown in \fig\ref{fig:finitea}.  
The statistical errors in
the $F_C$ curves are $\approx \pm 0.5$--1.4, and in the continuum extrapolation
$\approx \pm 2$; this is shown in the figure.

\subsection{Real time evolution (i): $\dpdt$}\la{sec:dphidt}

The factor $\dpdt$ is to be evaluated at the critical droplet value
$\avphi=\avphiC$.  It is easy enough to measure from numerical
simulations, but it can be also calculated {\em analytically}: from
the equations of motion \nr{eqm} we see that
\be
  \fr{\Delta\avphi}{\Delta t} = \fr{2}{N^3} 
	\sum_{x,a} \pi_a \phi_a + \mathcal O((\Delta t)^2)\,.
\la{psum}
\ee
Here all the fields and $\Delta t = (\Delta t)_{\rm cont.}/a$ 
are in dimensionless lattice units.  
In thermal equilibrium, the momenta $\pi_a(x)$ have Gaussian random
variable distribution (of width one in our normalization).  This
implies that $\< \pi_a(x)\pi_b(y)\> =
\delta_{a,b}\delta_{x,y}$, and any expectation values involving
products of $\pi$'s and $\phi$'s factorize: $\< f(\phi)
g(\pi)\> = \< f(\phi)\> \< g(\pi)\>$.  We need to determine $\< |
\Delta \avphi/ \Delta t | \>$ at the special value $\avphi = \avphiC$.  Using
this constraint, $\sum_{x,a} \phi_a^2(x) = N^3 \avphiC$,
we obtain
\be
  \bigg\< \bigg( \fr {\Delta\avphi}{\Delta t} \bigg)^2 
	\bigg\>_{\avphiC}
 =  \fr{4}{N^6} \sum_{x,a} \< \phi^2_a(x) \> =
    \fr{4\avphiC}{N^3}\,.
\ee
For any fixed $x$ the product $\pi_a(x)\phi_a(x)$ has a non-Gaussian
distribution, but the sum which appears in \eq\nr{psum} is Gaussian,
due to the global constraint.  Because Gaussian distributions satisfy
$\<x^2\>=\pi\<|x|\>^2/2$, we obtain the result
\be
  \bigg\< \bigg| \fr {\Delta\avphi}{\Delta t} 
	  \bigg| \bigg\>_{\avphiC}
  = \fr{1}{a} \, \sqrt{\fr{8 \avphiC}{\pi N^3}}\, ,
\la{fluxres}
\ee
where we have converted back to continuum time, to remind us of the
correct scaling in the continuum limit.  Since this result depends
only on the average distribution of momenta $\pi$, it is independent
of the magnitude of the noise in \eqs\nr{eqm}.  

\subsection{Real time evolution (ii): $\<\bd\>$}

The final contribution to the rate comes from 
$\<\bd\>=\< \delta_{\rm tunnel}/{N_{\rm crossings}}\>$.
This requires the evaluation of full real time trajectories through an
ensemble of critical droplets.  Following the procedure outlined in
Sec.~\nr{sec:mce}, we do this as follows:

\vspace{3mm}
\noindent (1) 
First, we choose an initial configuration $\phi^0_a(x)$ from a thermal
distribution, but with the order parameter restricted to a narrow
interval around the critical droplet value:
$|\avphi-\avphiC|<\epsilon/2$.  These are straightforward to generate
with either a standard Monte Carlo simulation with the restriction
built into the update, or by choosing them from a multicanonical run.  

\vspace{3mm}
\noindent (2)
We assign initial momenta $\pi^0_a(x)$ to this configuration, drawn
from a thermal distribution. In our case this means that we choose
$\pi_a(x)$ from a Gaussian distribution with width
$a^3\<\pi^2\> = 1$ (or $\< \pi^2 \> =1$ in lattice
units).

\vspace{3mm}
\noindent (3)
The configuration is then evolved in time until the order parameter $\avphi$
reaches $\avphiX{S}$ or $\avphiX{B}$, the symmetric or broken phase
``cut-off'' values.  We always use the timestep $\Delta t = 0.05 a$ in
our real time runs.  We then return to the initial configuration
$\phi^0$, invert the initial momenta $\pi^0 \rightarrow -\pi^0$,
and evolve the system again until we reach $\avphiX{S}$ or
$\avphiX{B}$.  This latter run is interpreted as a run backwards in
time; by gluing the backwards and forward half-trajectories together
at the starting point, we obtain a full trajectory $\avphi(S {\rm
~or~} B)\rightarrow\avphi(S {\rm ~or~} B)$.  If the trajectory
tunnels, i.e. if the backward and forward evolution ends are on different
sides of $\avphiC$, it contributes to the tunneling rate ($\delta_{\rm
tunnel} = 1$).  After counting the number of times the trajectory
crosses $\avphiC$, we obtain its contribution to $\<\bd\>$.

\vspace{3mm}

\begin{figure}[tb]
\centerline{
\epsfxsize=8cm\epsfbox{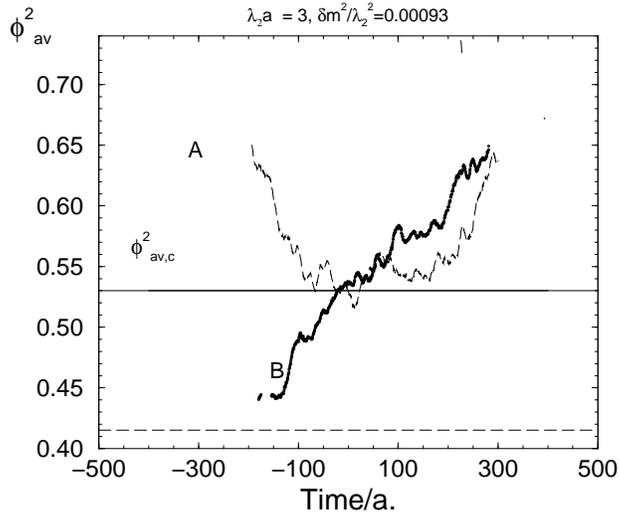}
}
\caption[a]{
Two trajectories measured from $\lambda_2 a = 3$, $80^3$ lattice, at
supercooling $\delta m^2/\lambda_2^2 = 0.00094$ and with noise magnitude
$\varepsilon^2 = 2\gamma \Delta t = 0.00125$.  The trajectories are
evaluated forward and backwards in time, starting from configurations
at $t=0$.  Trajectory A crosses $\avphiC$ 12 times, B 5 times;  
the contribution to $\bd$ from A is $0/12$, and from B $1/5$.  The
dashed horizontal line is the symmetric phase expectation value; and
$\avphi$ is shown in lattice units.}
\label{fig:trajectory}
\end{figure}

\begin{figure}[tb]
\centerline{
\epsfxsize=7cm\epsfbox{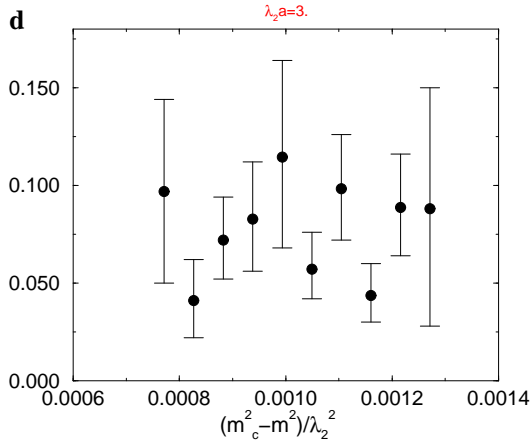}
}
\caption{The dynamical factor $\bd$ plotted against $\delta m^2$,
measured from $a\lambda_2=3$, $80^3$ lattice.}
\label{fig:dm2}
\end{figure}

An example of the trajectories is shown in \fig\ref{fig:trajectory}.
About 40\% of the trajectories we
evaluate are tunneling trajectories.  This means that the order
parameter resolves the critical droplet quite well, and configurations at
$\avphi=\avphiC$ indeed are a good sample of critical droplet
configurations (see the discussion in Sec.~\ref{sec:rateop}).  We
measure the order parameter after every timestep.  On the scale of the
plot the evolution of $\avphi$ looks rough, but at the level of
individual timesteps it is quite smooth.  This is due to the
relatively small amplitude of the noise in the equations of motion.

In Table \ref{tab:lats} we show the lattices and $\delta m^2$ values
where we evaluate the trajectories.  In \fig\ref{fig:dm2} we study how
$\bd$ depends on the degree of supercooling $\delta m^2$.  Remember
that small values of $\delta m^2$ correspond to large critical
droplets.  At each value of $\delta m^2$ we evaluated 35--83 full
trajectories.  The value of $\bd$ appears to be remarkably stable
throughout the range of $\delta m^2$ studied; the free energy of the
critical droplet varies by $\sim e^{100}$ over this range.  Naturally,
$\bd$ still depends on $\delta m^2$; for example, when $\delta
m^2\rightarrow 0$ one expects $\bd\rightarrow 0$, simply because
the larger the droplet is, the slower it evolves.
However, the dependence on $\delta m^2$ is not visible within our
statistical errors.  The large values of the errors are caused by the
large variation in the number of crossings between tunneling
trajectories: some trajectories have only a few crossings, whereas some
have more than 50. The errors of $\bd$ are still sufficiently small
for an accurate calculation of the the nucleation rate --- since the
rate is of order $e^{-100}$, a factor of 2 means little in the final
result.

The distribution of the trajectories as a function of the number of
crossings appears to decrease roughly exponentially as the number of
crossings increases.  However, when the number of crossings is small,
the trajectories with an even number of crossings --- which do not lead
to tunneling --- occur more frequently than those with an odd number of
crossings.  This is probably a consequence of the order parameter
fluctuations in the bulk phases, discussed in Sec.~\ref{sec:rateop}.

The lattice spacing dependence of $\bd$ is small: in addition to the
$a\lambda_2=3$ results in \fig\ref{fig:dm2}, we have measured it using
lattice spacings $a\lambda_2 = 2.5$ and $a\lambda_2=4$.  At
$a\lambda_2 = 2.5$, with a lattice of size $100^3$ and supercooling
$\delta m^2/\lambda_2^2 = 0.00094$, the result is $\<\bd\> =
0.049 \pm 0.022$; and at $a\lambda_2 = 4$, volume $60^3$ and
supercooling $\delta m^2/\lambda_2^2 = 0.001156$ the result is
$\<\bd\> = 0.064 \pm 0.018$.  Thus, no significant lattice
spacing effect is seen.  Note however that on extremely fine lattices,
$\< |\Delta \avphi/ \Delta t | \> \propto \sqrt{\avphi}\sim a^{-1/2}$ 
diverges when expressed in physical units (if we keep $\Delta t$ smaller
than the lattice spacing), and $\bd$ must correspondingly go to zero as
$a^{1/2}$.  

\subsection{Effect of the noise}

Let us study how the rate is affected by the magnitude of the
noise + damping terms, parametrized by $\gamma$ in the equations of
motion \nr{eqm}.  Since different levels of noise
correspond to different dynamical evolution (also in the continuum),
there is no reason why the results could not have significant
dependence on it.  As mentioned before,
$\dpdt$ is independent of the noise,
as is $P(\avphi)$, so it is sufficient to study how $\bd$
depends on $\gamma$.

What kind of behavior can we expect?  Naively, neglecting
the interactions and the mass ($\lambda_1 = \lambda_2 = m^2 = 0$) 
and the noise term $\xi$
in the equations of motion, 
the evolution of a mode of wave vector ${\bf k}$
obeys
$  
  \omega^2 - i\gamma\omega - k^2  = 0
$.  
If we have small $\gamma < 2k$, we see that $|\omega| =
k$ independent of the value of $\gamma$.  On the other hand,
if $\gamma > 2k$, the evolution becomes overdamped, and if
$\gamma$ is very large $\omega \approx i k^2/\gamma$.  Boldly
extrapolating these simple arguments to the physical evolution of the
critical droplet, we can expect that at small $\gamma$ the rate is
approximately constant as $\gamma$ increases,
whereas above a threshold value it should behave as $1/\gamma$.

\begin{figure}
\centerline{
\epsfxsize=8cm\epsfbox{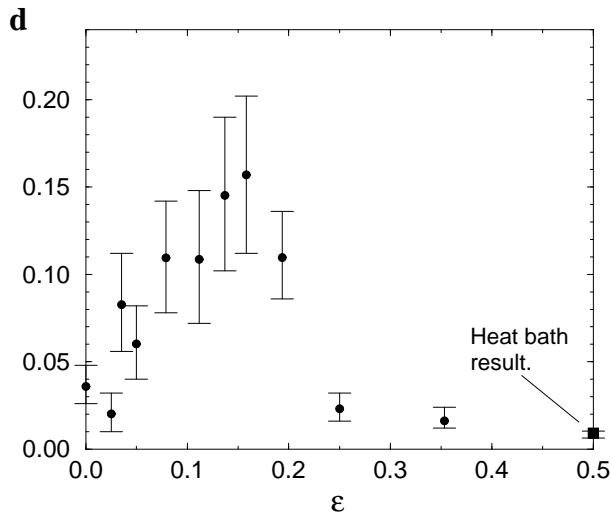}
}
\caption[a]{
The dependence of $\bd$ on the noise magnitude $\varepsilon = (1 -
e^{-2\gamma\Delta t})^{1/2}$, measured on an $a\lambda_2 = 3$, $80^3$
lattice at supercooling $\delta m^2/\lambda_2^2 = 0.00094$.  The point
at $\varepsilon = 0$ corresponds to Hamiltonian evolution, and the
``heat bath'' result to completely noisy evolution $\epsilon=1$,
shifted, for clarity, to smaller $\epsilon$.}
\label{fig:eps}
\end{figure} 

In \fig\ref{fig:eps} we show the measured value of $\bd$ against
$\varepsilon=(1-e^{-2\gamma\Delta t})^{1/2}$. Except for the points near
$\varepsilon=0$, we indeed observe roughly the expected behaviour, and,
surprisingly, even the threshold scale $\gamma \sim 2k \approx
2\pi/(\mathrm{droplet~size}) \approx 0.25$ (or $\varepsilon\sim 0.16$ in
\fig\ref{fig:eps}) is reproduced by the data.  At $\varepsilon
= 0$ the evolution is Hamiltonian, and, as discussed in
Sec.~\nr{sec:dyn}, the finite volume causes additional complications:
the growing/shrinking droplet releases/absorbs latent heat, which
rapidly equilibrates throughout the system.  Since the Hamiltonian
evolution conserves total energy, on a finite volume this causes
slight heating/cooling of the system.  This, in turn, tends to stabilize
the droplet, strongly reducing the tunneling
rate.  Indeed, in a small enough volume the critical droplet may
not decay at all.

The addition of the noise to the equations of motion thermalizes the
system effectively.  A natural amplitude for noise is $\gamma = 1/L$,
which thermalizes the system in the same timescale as 
waves propagate through it.  This corresponds to $\varepsilon=0.035$
in \fig\ref{fig:eps}.  Indeed, from this point up to $\varepsilon\sim
0.16$ $\bd$ does not vary much, and certainly the variation is not
significant for the final tunneling rate calculation.  Thus, we can
assume that in a very large volume, the Hamiltonian evolution
would also give a value for $\bd$ at this level or slightly larger.

\subsection{Nucleation rate}

\begin{figure}
\centerline{
\epsfxsize=7.5cm\epsfbox{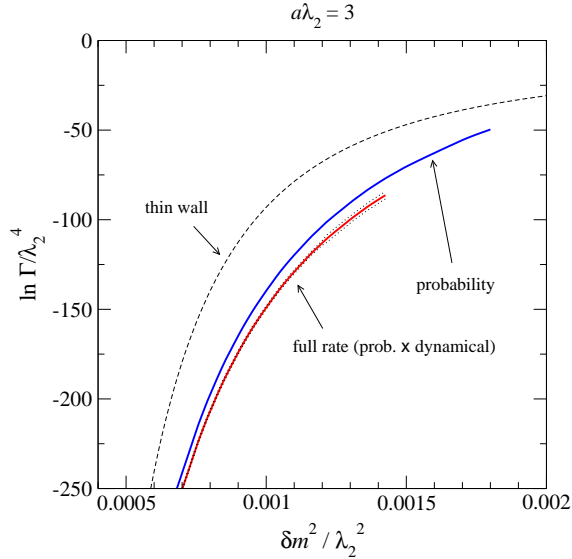}
}
\caption{The rate of droplet nucleation $\Gamma/\lambda_2^4$
from $a\lambda_2 = 3$ lattices. The full rate curve contains all
of the terms in \eq\nr{mcrate2}, whereas the ``probability'' curve contains
only the contribution $P_C^\epsilon/(V\lambda_2^3)$.  The thin
wall curve is calculated using the surface tension and the latent
heat measured from $a\lambda_2 = 3$ simulations.}
\label{fullrate3}
\end{figure}

Finally, let us pull together the ingredients discussed above and
calculate the rate $\Gamma/\lambda_2^4$ with \eq\nr{mcrate2}.  First,
note that since the dimension of $P_C^\epsilon$ is $[\avphiX{{\rm
lat}}]^{-1}=[a\avphiX{{\rm cont.}}]^{-1}$, it does not have a
good continuum limit by itself.  In order to cancel the $a^{-1}$ factor, it
is convenient to multiply $P_C^\epsilon/(V\lambda_2^3)$ by
$(a\lambda_2)$ before extrapolation, and correspondingly divide
$\dpdt\<\bd\>$ by it.  (This is not an issue for
$P(\avphiC)/P(\avphiX{{\rm SYM}})$ used in
Sec.~\ref{sec:probdist}, since it is dimensionless.)

In \fig\ref{fullrate3} we show the nucleation rate
$\Gamma/\lambda_2^4$ from $a\lambda_2 = 3$ lattices using $\gamma =
0.0125$, where we have the most extensive set of data.  In the
``probability'' curve we have set the product of the dynamical factors
$1/(2a\lambda_2)\dpdt\<\bd\>$ equal to $\lambda_2$ (this gives
correct dimensions), and only the probability $P_C^\epsilon$
contributes to the rate.  In the ``full rate'' curve we include 
the correct value of $\dpdt$ from \eq\nr{fluxres}, and $\bd$
from \fig\ref{fig:dm2}, by substituting it with its average 
and making a conservative error estimate, $\bd = 0.08\pm0.04$.
In the final error estimate we also take into account the errors of
$P_C^\epsilon$.

The inclusion of the dynamical factors reduce the rate by a factor of
$\sim 10^5$.  However, we emphasize that only the full rate is
independent on the choice of observables.  For example, if we simply
switch from an intensive to an extensive order parameter,
$P_C^{\epsilon}$, which is proportional to $1/\avphi$, would decrease
by a factor $\propto 1/V$; this is compensated by a corresponding
increase in $\dpdt$, so that the full rate remains invariant.

\begin{figure}
\centerline{
\epsfxsize=7.5cm\epsfbox{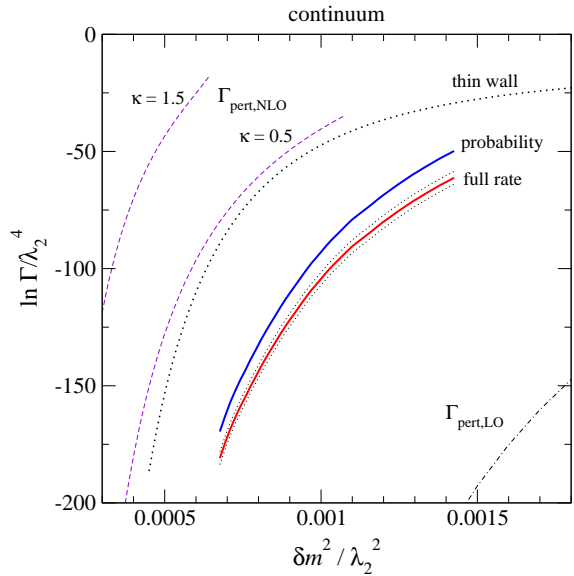}
}
\caption{The nucleation rate as in \fig\ref{fullrate3}, but 
extrapolated to the continuum limit.  For comparison, we also
show the rate calculated using the thin wall approximation and
leading order and next-to-leading order perturbation theory.
The curves labelled $\kappa=0.5$ and 1.5 show the scale sensitivity
of the NLO calculation, see Sec.~\ref{sec:pert}.}
\label{fullratecont}
\end{figure}

Next, let us consider the continuum limit extrapolation of the rate
$\Gamma/\lambda_a^4$.  
The extrapolation of the probabilistic factor 
$P_C^\epsilon(a\lambda_2)/(V\lambda_2^3)$ alone proceeds as in
\fig\ref{fig:finitea}; we fit an ansatz of form $c_0 + c_2 a^3$
independently at each $\delta m^2/\lambda_2^2$ to the lattice results
from $a\lambda_2 = 2$, 2.5 and 3.  The resulting curve is shown in
\fig\ref{fullratecont}.  The remaining contribution to the rate is
given by the dynamical factor $(a\lambda_2)^{-1} \dpdt\<\bd\>$, which
we assume to be constant as the lattice spacing is changed but the
physical volume is kept constant.  This quantity must have a continuum
limit unless the dynamics posess some UV pathology.
The simulation results from
different lattice spacings do not show any lattice spacing dependence
within statistical errors; however, errors are too large for a real
continuum extrapolation.  

The result of the extrapolation is shown in \fig\ref{fullratecont}.
The errors of the log of the rate are $\approx \pm 4$, including the
errors coming from the extrapolation of $P_C^\epsilon$ and the
estimated error from the extrapolation of $\bd$.  The error in
$P_C^\epsilon$ dominates the uncertainty.

\subsection{Droplet cross-sections}\la{sec:cross}

\begin{figure}[tb]
\centerline{
\epsfxsize=13cm\epsfbox{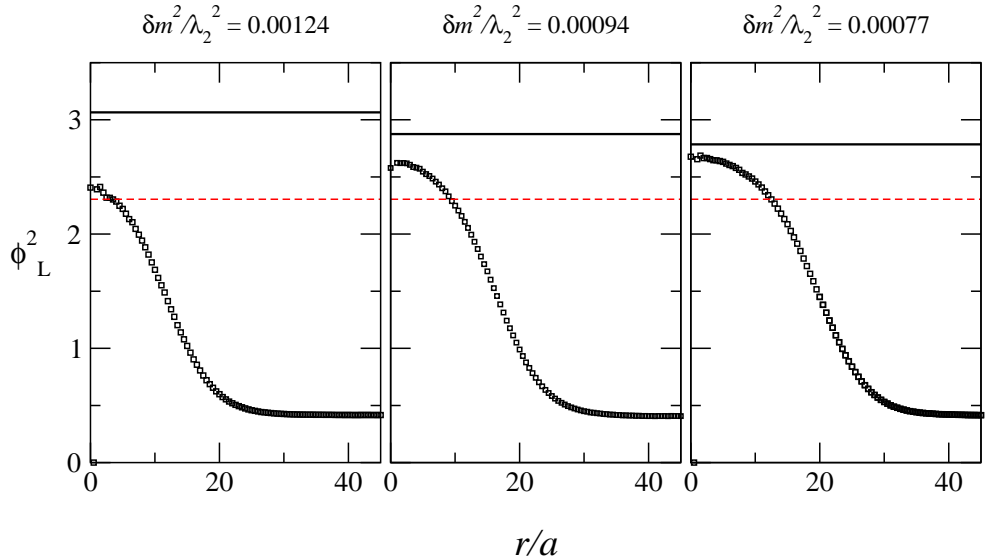}
}
\caption[a]{The droplet cross-sections at 3 different
supecoolings from
$a\lambda_2 = 3$, volume $80^3$ lattices.  The expectation value
$\<\phi_L^2\>$ is shown in lattice units.  The dotted line shows the
broken phase $\<\phi_L^2\>$ at the critical value of $m^2$, the solid lines
at the supercooling $\delta m^2$.  Only at the smallest supercooling
does the core of the droplet approach the broken phase expectation
value.}
\label{wallprofiles}
\end{figure}

It is illuminating to take a closer look at the structure of the
droplet configurations.  Since the droplets are large objects in terms
of correlation lengths, it is possible to perform suitable averaging
over lattice configurations in order to measure the size and shape of
critical bubbles at fixed supercooling.  We do the averaging as
follows:
\begin{itemize}\itemsep=0cm

\item Generate a number of configurations with the
constrained order parameter $|\avphi-\avphiC|<\epsilon$, where
$\epsilon$ is a small number.  These configurations contain a droplet
of a fixed volume, as discussed in Sec.~\nr{sec:droplet}.

\item Determine the center of the droplet.  We do this by averaging the
order parameter over $y,z$ coordinates, and taking the lowest non-trivial
Fourer mode 
$
  A = \sum_x \phi^2(x) e^{i 2\pi x /L}\,.
$
The value $x_0 = L/\pi\, \arctan(\mathrm{Im\,} A/\mathrm{Re\,} A)$
gives now the point around which the configuration is maximally
symmetric.  This is repeated for $y$ and $z$.

\item Shift the origin to the center of the droplet, and determine
the average order parameter as a function of the radius, $\phi^2(r)$. 
Averaging over the configurations we obtain the average droplet 
cross-section $\<\phi^2(r)\>$.

\end{itemize}
We determined the droplet cross-sections for several different $m^{2}$
at $a\lambda_2 = 3$ lattices of size $80^3$, and in
\fig\ref{wallprofiles} we show the results for supercooling $\delta
m^2/\lambda_2^3 = 0.00124$, 0.00094 and 0.00077.  From
\fig\ref{fig:finitesize} we see that these correspond to droplet free
energies from 75 to 180.  The droplet wall is very thick, and
especially for smaller droplets there is no broken phase core to speak
of.%
\footnote{%
	One should be a little cautious of interpreting the interface
	width shown in Fig.~\ref{wallprofiles}, however, since
	an interface in 3 dimensions generically exhibits 
	logarithmically large roughening.  Locally the interface
	may be thinner.%
	}  
Thus, the thin wall approximation for the nucleation rate cannot
be expected to work well.  The expectation value $\<\avphi\>$ is
larger at the center of the droplet than the broken phase expectation
value at $m_c^2$; this is naturally because the broken
phase expectation value increases rapidly as $m^2$ is decreased.  Only
for the largest droplet (largest $m^2$) does the core approach the
broken phase expectation value.

One can note that although the bulk of the droplets fits well
on the lattice, at a smaller volume the exponentially decreasing tail
of the fields outside the droplet wall would feel the lattice size.
This decreases the free energy of the droplets, and it is no surprise
that the $60^3$ lattice gives smaller droplet free energy at large
droplets, see \fig\ref{fig:finitesize}.  

\section{Comparison with other nucleation rate calculations}\la{sec:compare}

\subsection{Thin wall approximation}

The simplest semi-analytical estimate for the nucleation rate
can be obtained using the thin wall approximation for the
critical droplet free energy \eq\nr{fdroplet}:
\be
   F_{TW} = \fr{16\pi\sigma^3}{3(\delta m^2\ell)^2} \, ,
\ee
where we use lattice determinations of the surface tension $\sigma$
and latent heat $\ell$.  Thus, this method is not fully 
analytical; however, normally the determination of
these values is an order of magnitude easier task than the full
nucleation rate computation, and the thin wall
approximation has often been used to include non-perturbative
input in the nucleation rate calculations (see, for example,
\cite{Iwasaki}).

The free energy is converted to a rate by multiplying it with suitable
mass scale
\be
  \Gamma = m^4 \exp -F_{TW}\,.  \la{ftorate}
\ee
We use here the mass $m=0.05 \lambda_2$, which is close to the
symmetric phase perturbative mass; the precise value is not
significant.  The thin wall approximation assumes that the thickness
of the droplet wall is negligible and that there are no contributions
to the free energy due to the curvature of the walls.  
It becomes valid in the limit where the
droplet radius is much larger than the thickness of the droplet wall;
this condition is not usually well met in practice (see
Sec.~\ref{sec:cross}).

In \fig\ref{fullrate3} we show the thin wall result using the
values of $\sigma$ and $\ell$ obtained from $a\lambda_2=3$
simulations.  In the range of interest the thin wall rate is higher by
$\sim e^{50}$.  Large as the difference is, the thin-wall
approximation is not as bad as one may first think: in many physical
processes the system is cooled down at a very slow (constant) rate, and
thus, the nucleation rate is changing with time: $d \Gamma/dt > 0$.
Since $\Gamma$ depends exponentially on $\delta m^2$, the transition
occurs very rapidly when the nucleation rate reaches some critical
value, typically $\Gamma_c \times l^3 (t-t_c) \sim 1$, where $l$ is
some relevant length scale.

Let us take $\Gamma/\lambda_2^4 = e^{-100}$ as our reference value.
The lattice results from $\lambda_2=3$ give a supercooling value
$\delta m^2/\lambda_2^2 = 0.00132(2)$, whereas the thin wall result is
$0.00096(4)$, which is ``only'' $\sim 30$\% smaller than the correct value.
The errors of the thin wall results come from the errors of the
determination of $\sigma$ and $\ell$.

In \fig\ref{fullratecont} we compare the lattice results to the thin
wall ones at the continuum limit.  Using again the reference value
of $\Gamma_c/\lambda_2^4 = e^{-100}$, the lattice result for the
supercooling is $\delta m^2/\lambda_2^2 = 0.00094(3)$, and for the 
thin wall calculation $0.00063(7)$, again a 30\% difference.
Thus, the thin wall calculation gives a rather good order of magnitude 
estimate for the nucleation rate.  This was also seen in the
SU(2)-Higgs calculation in Ref.~\cite{ewbubble}. 

\subsection{Perturbation theory}\la{sec:pert}

In the broken phase of the cubic anisotropy model (and with our choice
of $\lf$) the field component which acquires a non-zero expectation
value ($\phi_1$, say) is much lighter than the other field component
($\phi_2$).  Due to this mass hierarchy, it turns out that we can
calculate an effective potential (or action) for the $\phi_1$ field
by perturbatively integrating over $\phi_2$.  This effective potential
can be used to calculate the leading contributions to the critical
droplet free energy by finding the classical ``bounce'' solution to
the equations of motion \cite{EIKR,Turok,Dine}.  More concretely,
assuming that the droplet is spherically symmetric and centered at
$r=0$, we want to find a configuration $v(r)$ which is a saddle point
of the action
\be
   S(m^2) = 4\pi \int_0^\infty dr\, r^2 \,\left[ 
	\fr12 (\partial_r v)^2 + V(m^2; v)\right]\,.
\la{bounce}
\ee
Here $v$ is the expectation value of the scalar field.
The boundary conditions are $v(\infty) = 0$, $\partial_r
v(0) = 0$.

The dimensionless expansion parameter is proportional to
the ratio of the two couplings $\lf$.  This implies that the
convergence becomes better the stronger the first order transition is
(see Sec.~\ref{sec:cubic}).   For our parameters the
expansion parameter is $1/8 \ll 1$, offering formally a good 
convergence parameter.

Without loss of generality we can choose the transition to happen
in the direction of the $\phi_1$ field, and we shift the fields
$(\phi_1,\phi_2)\rightarrow(v+\phi_1,\phi_2)$.  The tree (mean
field) level potential is
\be
  V_0(v) = \fr12 m^2 v^2 + \fr1{4!} \lambda_1 v^4 
\la{v0}
\ee
and the 1-loop potential is
\be
  V_1(v) = - \fr1{12\pi} (m_1^3 + m_2^3)\,,
\la{v1}
\ee
where 
\be
  m_1^2  =  m^2 + \fr12 \lambda_1 v^2\,, ~~~~~~~~~
  m_2^2  =  m^2 + \fr12 \lambda_2 v^2\,.
\la{m12}
\ee
In the broken phase, for $m^2$ close to the transition value, 
inspection of the potential gives the following power counting rules:
\be
  v^2\sim\lambda_2^3/\lambda_1^2 
  ~~~~ \mbox{and} ~~~~~
  m^2\sim\lambda_2^3/\lambda_1\,.
\la{eq:count}
\ee
The leading contribution from the 1-loop term $m_2^3$ is
of the same order as the tree level terms; it is just this third order
term which makes the transition first order.  On the other
hand, the contribution from $m_1^3$ is suppressed by a factor
$\propto(\lf)^{3/2}$ in comparison to the leading order.  This is
actually subleading when compared with the leading 2-loop
contributions; furthermore, at this order one also gets contributions
from the resummation of the $\phi_1$ propagator, which render $m_1$
formally imaginary (easily seen by considering $V_1''(v)$,
and equating this to $m_1^2$).  It actually 
turns out that up to next-to-leading order in $\lf$ we do not
have to consider the fluctuations of $\phi_1$ at all, justifying
the use of the classical ``bounce'' solution at this order.

At the leading order we can also neglect $m^2$ in the expression
for $m_2^2$, and write the potential as
\be
  V_{\rm LO}(v) = V_0(v) - 
	\fr{1}{24\sqrt{2}\pi} \lambda_2^{3/2} v^3\,.
\ee
This potential gives a very strong first order transition, much
stronger than the lattice calculations indicate, as can be seen by
comparing the broken phase values $v^2$ and the surface tension
$\sigma$ in Table \ref{tab:pert}.  We calculate the perturbative
surface tension from the integral
\be
  \sigma = \int_{v_{\rm symm.}}^{v_{\rm broken}} dv \sqrt{2 V(v)}\,.
\ee
The droplet free energy (at some value of supercooling) can be solved
from \eq\nr{bounce} numerically using the well-known
undershoot/overshoot method \cite{numrec}.  (It can be solved
analytically in some limiting cases, see, for example, \cite{EIKR}.)
As in the thin wall droplet case, we convert the droplet free energy
to the nucleation rate using $\Gamma = m^4
\exp(-F)$, where we use the mass scale $m=0.05\lambda_2$ (the result is
insensitive to the precise value of $m$).  Solving for the
value of supercooling which gives the reference rate 
$\Gamma/\lambda_2^4 = \exp -100$, 
we obtain the supercooling $\delta m^2 = 0.00234 \lambda_2^2$, which
is more than 100\% larger than the result from the lattice simulations,
\fig\ref{fullratecont} and Table~\ref{tab:pert}.

The next-to-leading order corrections to the potential are suppressed
relative to the leading terms by $\lf$.  They arise from 
$m_2^3$ in \eq\nr{v1}, and from the leading two-loop
contribution:
\be
  V_{2}(v)  =  V_0(v) - \fr{1}{12\pi} m_2^3 
                  -  \fr{v^2 \lambda_2^2 }{4(4\pi)^2} \left[ 
                     \fr{1}{2} \log\fr{\mu^2}{2 \lambda_2 v^2} 
		+ \fr{1}{2}\right]
\la{v2}
\ee
The last term above is the two-loop contribution; it arises from the
logarithmically divergent ``sunset'' diagram with two $\phi_2$
-propagators and one $\phi_1$-propagator, see \fig\ref{pgraphs}.
At this order we can neglect the masses from the propagators in the
2-loop diagram, and also the other 2-loop contributions (sunsets and figure-8
diagrams), which are suppressed by $(\lf)^{3/2}$ or more.  

\begin{figure}[tb]

\centerline{\epsfxsize=7.5cm\epsfbox{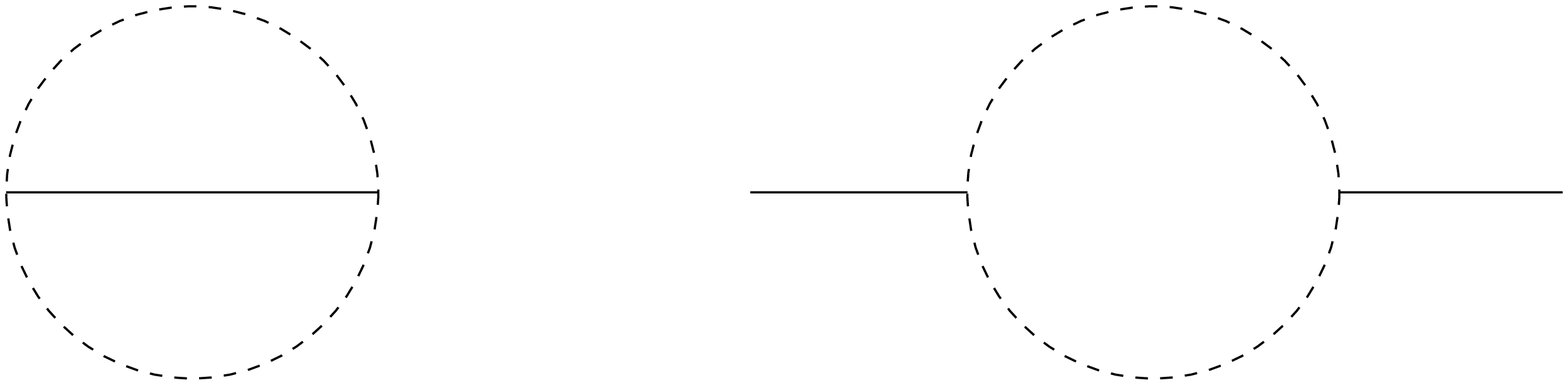}}
\caption[a]{Left: 2-loop ``sunset'' diagram used in the effective
potential calculation.  Right: the diagram used in the calculation
of the wave function correction.  The dotted lines are the ``heavy''
$\phi_2$ -propagators, and the solid lines correspond to the 
``light'' $\phi_1$ propagators.  The coupling constant at the
vertices is $\lambda_2 v$.}
\la{pgraphs}
\end{figure}

\begin{table}
\centerline{
\begin{tabular}{|ll|l|l|l|}
\hline
           &    &$v_{\rm broken}^2/\lambda_2$&$\sigma/\lambda_2^2$&
		$\delta m^2/\lambda_2^2$ \\
\hline
\multicolumn{2}{|l|}{$V_{\rm LO}$} & 0.81   & 0.0124    & 0.0023      \\
$V_{\rm NLO}$&$\kappa=0.5$         & 0.44   & 0.0034    & 0.00058     \\
             &$\kappa=1.5$         & 0.40   & 0.0027    & 0.00033     \\
\multicolumn{2}{|l|}{lattice}      & 0.49(1)& 0.0050(2) & 0.00094(3) \\
\multicolumn{2}{|l|}{thin wall}    & -      & -         & 0.00063(7) \\
\hline
\end{tabular}}
\caption[a]{Comparison of the perturbative, lattice and thin wall
results.  The supercooling has been evaluated at the point where
the rate $\Gamma/\lambda_2^4 = \exp(-100)$.
\la{tab:pert}}
\end{table}

In \eq\nr{v2} $\mu$ is the scale factor arising from dimensional
regularization in $d=3-2\epsilon$ dimensions.  As usual, a logarithmic
divergence has been absorbed in a mass counterterm, which defines
the scale dependent running mass
\be
  m_R^2(\mu) = \fr{\lambda_2^2}{2(4\pi)^2} \log\fr{\mu}{\Lambda}\,.
\la{mr}
\ee
We substitute $m^2$ with $m_R^2$ everywhere where it
appears in the potential.

The dependence of $V(v)$ on the scale $\mu$ is formally of higher
order than $\lf$; thus, we are free to choose $\mu$ so that the effect
of the large logarithms in the potential are minimized.  A common
prescription is to fix $\mu$ so that the logarithm vanishes in the
broken phase.  Naturally, this does not suppress the logarithms in the
symmetric phase at all.  A better overall cancellation of the logs is
achieved by using a $v$-dependent 
scale $\mu = 2\kappa \sqrt{\lambda_2 v^2/2}$, 
where $\kappa$ is a constant of order unity.  However, this simple
prescription is not complete, as discussed in Sec.~5 of Ref.~\cite{FKRS}.
Simplifying the renormalization arguments in the above reference a bit,
this follows from the fact that the
value of the effective potential is not a physical quantity, only the
differences $V(v_1)-V(v_2)$ are, and the absolute value of the
potential can have unphysical $\mu$-dependence.  (Indeed, in
dimensional regularization the value of the potential also at $v=0$
depends on the scale $\mu$.)  Thus, if we change the renormalization
point $\mu$ as we change $v$, the ``normalization'' of the potential
changes, and it is not possible to directly compare $V(v_1,\mu(v_1))$
and $V(v_2,\mu(v_2))$ if $v_1$ and $v_2$ are far away from each other.

The most straightforward solution to this problem is to consider the
slope of the potential $dV(v)/dv$ instead
of the potential itself.   With this quantity we can write
a renormalization group improved expression for the next-to-leading
order effective potential
\cite{FKRS}:
\be
  V_{\rm NLO}(v') = \int_0^{v'} dv
  \left.\fr{dV_2(v,\mu)}{dv}\right|_{\displaystyle
       \mu=2\kappa\sqrt{\lambda_2 v^2/2}}\,.
\la{vimpr}
\ee
The derivative in the above expression is to be evaluated at fixed
$\mu$, and only after taking the derivative $\mu$ is set to its
optimized value.  

In Table \ref{tab:pert} we show $v^2_{\rm broken}$ and $\sigma$,
calculated from potential \nr{vimpr}, using values $\kappa=0.5$ and
1.5.  The transition is weaker than the results from the lattice
simulations; the broken phase expectation value of $v^2_{\rm broken}$
is 10-20\% smaller, but the surface tension is around 40\% smaller.
Likewise, the supercooling $\delta m^2$ at rate $\Gamma =
\lambda_2^4\exp(-100)$ is around 50\% smaller than the lattice
results.  This discrepancy is of similar magnitude to the difference
between perturbative and 2-loop results in the nucleation rate in
SU(2)-Higgs theory \cite{ewbubble}; however, in contrast to the cubic
anisotropy model, the supercooling at a fixed value of the rate was
seen to be {\em larger} in the perturbative analysis than on the
lattice.

The scale dependence of the NLO nucleation rate is quite large in 
Table~\ref{tab:pert}: the value of the
supercooling increases by 75\% when $\kappa$ is decreased from 1.5 to
0.5.  This suggests that higher orders would affect the results
significantly --- this is, of course, already indicated by the fact
that the NLO results do not agree well with the leading order results,
or with the lattice.  The large
scale dependence is not a feature of the procedure to set $\mu$ used
in \eq\nr{vimpr}:  we can, for example, use the 2-loop potential 
\nr{v2} and just fix the scale $\mu$ so that the 2-loop contribution
vanishes in the broken phase.  If we vary $\mu$ by a factor of $\sim
50$\%, the strength of the transition changes about as much as when we
vary $\kappa$ in \eq\nr{vimpr}.  Moreover, this method tends to give a
somewhat weaker transition than \eq\nr{vimpr}.

In principle, finding the extremal critical droplet solution of the
bounce action \eq\nr{bounce} with the NLO potential \nr{vimpr} does
{\em not} give all of the $\mathcal{O}(\lf)$ contributions.  At this
order one should take into account also the {\em wave function
correction\,} to the kinetic energy in the bounce action by modifying
the second derivative term:
\be
  (\nabla v)^2 \rightarrow (1+Z_{\phi}) (\nabla v)^2 \,.
\ee
The leading contribution to $Z_{\phi}$ can be calculated by
evaluating the diagram on the right in \fig\ref{pgraphs} and expanding
the result to second order in the external momentum (see, for example,
Ref.~\cite{BBFH}).  This calculation is justified because
$v$ (or $\phi_1$) varies on length scales $\sim m_1^{-1}$, which
is much longer than the inverse mass of $\phi_2$, $m_2^{-1}$.  
The result of the calculation is
\be
  Z_{\phi} = -\fr{\lambda_2^2 v^2}{192 \pi m_2^3}\,,
\la{Z}
\ee
which is indeed $\propto\lf$ according to the power counting
rules, Eq.~(\ref{eq:count}).  
However, the inclusion of the wave function correction
modifies the results by less than 1\%, which is a negligible effect
considering the accuracy of the NLO calculation.  This is in contrast
to the SU(2) gauge + Higgs theory, where the wave function normalization was
seen to improve the results dramatically.  The reason for the
smaller effect in this case is the very small numerical multiplicative
factor in \eq\nr{Z}.

We can conclude that the NLO perturbative analysis at $\lf=1/8$
describes the phase transition qualitatively correctly, but the
physical quantities can be a factor of 2--3 off the correct ones.
Naturally, for $\lf$ closer to unity the NLO analysis would be even
less accurate.  The overall accuracy is comparable to the perturbative
treatment of the SU(2) gauge + Higgs theory \cite{ewbubble}.  Trying
to improve the perturbative treatment would require the calculation of
contributions proportional to $(\lf)^{3/2}$.  
At this order the light $\phi_1$ loops start to
contribute, and, as mentioned above, $\phi_1$ propagators require
resummation, which makes $m_1$ (and the 1-loop contribution in
\nr{v1}) formally imaginary in the region where the second derivative
of the effective potential is negative.  Thus, it is not consistent to
solve the effective potential in a constant background field as above;
the correct treatment would require the calculation of the fluctuation
determinant around a classical droplet solution.  The simultaneous
consistent (as a power series of $\lf$, say) evaluation of the
fluctuation determinant and the effective potential is a very
non-trivial problem, and we do not attempt it here.\footnote{The
fluctuation determinant has been recently performed by several
authors \cite{MunsterRotsch,StrumiaTetradis,Baacke,Parnachev} in various
physical systems.  However, these studies did not consider the
simultaneous order-by-order evaluation of the perturbative effective
potential and the fluctuation determinant.}

\subsection{Comparison with the coarse-grained effective action approach
}

The nucleation rate in the cubic anisotropy model was studied by
Strumia and Tetradis using a coarse-graining approach
\cite{StrumiaTetradis}.  
First one calculates a coarse-grained effective action, valid for
momentum scales smaller than a chosen scale $k$, where $k\lsim m$,
some physically relevant mass scale (for example, the mass of the
light $\phi_1$ field in the broken phase, if the transition happens
along the direction of $\phi_1$).  Formally, the effective action is
found by integrating out momenta $p>k$.  In the second step, the
nucleation rate is calculated using the effective action and Langer's
method: one finds the classical bounce solution (as in
Sec.~\ref{sec:pert}) with action $S_k$, around which the fluctuation
determinant $A_k$ can be evaluated.  The scale $k$ acts automatically
as the ultraviolet regulator for the fluctuation determinant.  The
full rate is then estimated to be
\be
   \Gamma = A_k e^{-S_k}\, .
\ee
Since the physics does not care about the artificial scale $k$,
the final result should be independent of $k$ if both halves of the
calculation are under control.

The authors of Ref.~\cite{StrumiaTetradis} find the effective action
using the renormalization group flow method of Wetterich
\cite{Wetterich}: they start from the bare action, \eq\nr{energy},
defined at some ultraviolet scale $k_0 \gg k$, and use the flow
equations to evolve the action down to scale $k$.  During the
evolution the action develops so that the resulting effective action
has a first order transition already at ``tree level''.  It still
contains both of the original fields.  The evolution generates
very complicated (non-local) effective actions, and in practical
calculations the actions have to be truncated to a simple ansatz.  For
details, see Ref.~\cite{StrumiaTetradis} and references therein.

In Ref.~\cite{StrumiaTetradis} the nucleation rate is calculated at
coupling constant ratios $\lf = 0.3$ and $0.15$ (in their notation,
$\lambda_{k_0}/g_{k_0} = 0.1$ and 0.05, defined at the ultraviolet
scale $k_0$.).  The latter is close enough of the value we use here
$\lf=1/8$ to make qualitative comparison possible.  The results show a
very strong dependence on the coarse graining scale $k$: for a fixed
supercooling, the nucleation rate varies from $\ln\Gamma/m_1^4 \approx
-250 $ to $ -130$, when $k/m_1$ is changed from 0.6 to 0.9.  Here
$m_1$ is the mass of the light $\phi_1$ field in the broken phase.
This implies that accuracy is lost during the calculation.
Furthermore, the leading contribution to the rate does not come from
$\exp(-S_k)$, but from the fluctuation determinant $A_k$.  This is
partly caused by the fact that the mass of the $\phi_2$ field in the
broken phase is much heavier than the coarse graining scale $k$, and
gives a large contribution to the determinant.  Indeed, one can argue
that a coarse grained effective action where the heavy field is
completely integrated out, as was done in Sec.~\ref{sec:pert}, may
offer a better starting point for the Langer method.\footnote{This
procedure has been suggested by Weinberg
\cite{Weinberg}: one obtains an effective action for the light fields
by integrating over heavy fields perturbatively.  Only the light field
fluctuation determinant is evaluated.  However, as discussed in
Sec.~\ref{sec:pert}, this is very difficult to implement as a 
correct order-by-order perturbative expansion.}

We should emphasize that the approach just described makes two separate
approximations.  The first is the truncation of the infrared effective
action, made in the the renormalization group flow part of the
calculation as the coarse graining scale is run down to $k$.  This
approximation means that some ultraviolet physics is lost or incorrectly
incorporated in the coarse grained effective theory.  The second
approximation is in the calculation of the nucleation rate within the
coarse grained theory itself; computing Gaussian fluctuations about a
classical saddle point corresponds to carrying perturbation theory to
one loop, neglecting interactions between infrared fluctuations.  This
is clearly not warranted if the infrared behavior is strongly coupled,
for instance.

Since the dominant contribution does not come from the saddle point
but from the fluctuation determinant, Langer's method as a saddle
point expansion fails in this calculation.  However,
our lattice simulations clearly indicate that a well-defined
saddle point exists in the phase space.  While the simulations make no
distinction between the classical solution and the fluctuation
determinant (everything is contained in the droplet free energy), the
configurations at the saddle point consist of well-defined critical
droplets.  Thus, the physical picture given by Langer's theory 
is valid; the problem in (semi)analytical calculations is to find
the correct effective description.

\section{Conclusion}\la{sec:conclusion}

In this paper we have used a novel technique for studying the critical
droplet nucleation rate in first order phase transitions using
non-perturbative lattice simulations.  The method is readily
applicable to exponentially small nucleation rates, which are often of
relevance in physically interesting transitions.  The technique consists of
two separate stages: (1) the Monte Carlo evaluation of the nucleation
barrier, using multicanonical methods, and (2) the correct treatment
of the microscopic dynamics of the nucleation process with real time
simulations.  The first stage can be considered as a generalization of
Langer's theory of nucleation \cite{Langer}: we have substituted the
approximate saddle point expansion with a nonperturbative Monte Carlo
calculation.  The second step goes partly beyond Langer's formalism.
This method is readily applicable to any theory which is amenable to
the lattice treatment in the first place, and, within the context of
the lattice, the technique is exact up to exponentially small and in
practice negligible corrections.  This method has also been applied to
the nucleation rate in SU(2)-Higgs theory \cite{ewbubble}.
Due to the multicanonical methods used, the statistical errors in 
the final answer tend to be automatically small --- usually much smaller
than any uncertainty in analytical approaches.

In this work we have applied the method to the phase transition in the
cubic anisotropy model, which is a formally simple field theory with
two scalar fields and a radiatively induced first order phase
transition, the strength of which can be adjusted continuously.
Because of the formal simplicity the theory is well suited for both
analytical and numerical analyses.  Since the transition is radiatively
induced, Langer's theory has to be applied with great care in analytical
calculations.

We compare the nucleation rate obtained from the lattice simulation to
customary analytical or semianalytical approaches.  We find that the
straightforward application of perturbation theory up to
next-to-leading order in the relevant expansion parameter is not
accurate: for a fixed value of the nucleation rate, the corresponding
supercooling is roughly a factor 2 off the correct value, even though
the scalar field expectation value is correct to within 15\%.  To go
beyond the next-to-leading order would require the evaluation of the
fluctuation determinant, and we did not attempt it in this work. On
the other hand, a thin wall estimate, using non-perturbative input, is
off only by 30\%.  The nucleation rate for a fixed supercooling is off
even more than these numbers.  This behaviour is strikingly similar to
that observed in SU(2)-Higgs theory.  The nucleation rate in the cubic
anisotropy model has also been studied in Ref.~\cite{StrumiaTetradis},
using an approximate coarse grained effective potential as a starting
point for Langer's procedure.  The results were seen to depend very
strongly on the coarse graining scale, making quantitative comparison
impossible.  

The results show that, for a rough-and-ready estimate of the
nucleation rate, the thin wall approximation is acceptable, provided
that one uses non-perturbatively determined surface tension and latent
heat as an input.  These are much easier to determine on the lattice
than the full nucleation rate.  On the other hand, the purely
perturbative analysis shows weak convergence.  If high accuracy
is required, one has to resort to numerical evaluation.

\subsection*{Acknowlegments}

We thank Dietrich B{\"o}deker and Mikko Laine for many useful
discussions.
K.R. acknowledges partial support from EU TMR grant FMRX-CT97-0122.
This work 
was partially supported by the DOE under contract DE-FGO3-96-ER40956.

\appendix

\section{Renormalization of the lattice theory}\la{sec:app}

This appendix details the matching calculation which eliminates $O(a^2)$
errors in the lattice theory.  This is done by computing a set of
correlation functions in the lattice and continuum theories, at one and
two loops, and adjusting the lattice couplings so the results coincide.
The required correlation functions are the two and four point functions
at zero momentum, the leading momentum dependence of the two point
function, an insertion of $\< \phi^2 \>$ on a zero momentum line, 
and the vacuum value of the $\< \phi^2 \>$ operator.  It is
necessary only to find the difference between lattice and continuum
values, to perform the matching; this difference is IR finite for all
graphs we need.  Each loop order eliminates errors at one higher power
of $a$ because graphs become more UV convergent by one power per loop
order in this 3 dimensional theory.  For a more thorough discussion of
how the matching calculation works, see \cite{Oapaper}.

\subsection{results}

\begin{figure}[t]
\centerline{\epsfxsize=6.5cm\epsfbox{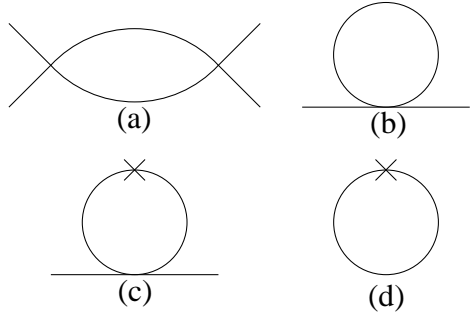}}
\caption{\label{one_loop_graphs} The one loop graphs needed for the
renormalization.  A cross represents a $\phi^2$ insertion.}
\end{figure}

A one loop lattice-continuum matching (renormalization) 
calculation will determine the $O(a)$
contributions to the quantities $Z_{\phi}$, $\delta \lambda$, and $Z_m$,
and will find the $O(1/a)$ contributions to $\delta m^2$
and $\delta \< \phi^2 \>$.  The required graphs are shown in
\fig\ref{one_loop_graphs}.  Evaluating the graphs requires choosing a
lattice Laplacian.  We consider two choices; the unimproved Laplacian 
\be
 \nabla^2_{\rm U} \phi(x) = -6 \phi(x) + \sum_i (\phi(x+i)+\phi(x-i)) 
	\, , \la{unimp_Laplace}
\ee
and the improved Laplacian, which we actually use:
\be
\nabla^2_{\rm I} \phi(x) = - \frac{15}{2} \phi(x) + \frac{4}{3} \sum_i
	\Big( \phi(x+i) + \phi(x-i) \Big) - \frac{1}{12} \sum_i \Big( 
	\phi(x+2i) + \phi(x-2i) \Big) \, . 
\ee
We write the Fourier transforms of these choices as 
\ba
\tilde{k}^2_{\rm U} & = & \sum_i ( 2 - 2 \cos(k_i) ) \, , \\
\tilde{k}^2_{\rm I} & = & \sum_i \left( \frac{5}{2} - \frac{8}{3} 
	\cos(k_i) + \frac{1}{6} \cos(2 k_i) \right) \, .
\ea
The evaluation of the one loop graphs requires two integrals:
\ba
\frac{\Sigma}{4 \pi} & \equiv & \int_{BZ} \frac{d^3 k}{(2\pi)^3} 
	\frac{1}{\tilde{k}^2} \, , \\
\frac{\xi}{4 \pi} & \equiv & \int_{BZ} \frac{d^3 k}{(2\pi)^3} 
	\frac{1}{(\tilde{k}^2)^2} - \int_{\Re^3} \frac{d^3 k}{(2\pi)^3} 
	\frac{1}{k^4} \, .
\ea
Here we use the shorthand $BZ$ to mean $k$ lies within the Brillouin
zone, meaning each $k_i \in [-\pi,\pi]$.  The integrals which determine
$\xi$ are each IR singular and some regulation is implied, for instance
adding $m^2$ to both $k^2$ and $\tilde{k}^2$ and taking the limit as
$m^2 \rightarrow 0$.  The numerical values of the integrals are
\ba
 \Sigma_{\rm U} & = & 3.17591153562522 \, , \qquad \:
 \Sigma_{\rm I}   =   2.75238391130752 \, , \nonumber \\
 \xi_{\rm U}    & = & 0.152859324966101 \, , \qquad
 \xi_{\rm I}      =  -0.083647053040968 \, .
\ea
Note that the sign of $\xi$ depends on whether we use an improved
lattice Laplacian.  This is possible because $\xi$ represents the
difference of a graph between lattice and continuum theories.  The
lattice contribution is larger inside the Brillouin zone, but the
continuum integral receives contributions from outside the zone as well;
the sign depends on which effect is larger.

At one loop the renormalizations are
\ba
\delta \lambda_{1,1l} & = & \left( \frac{3}{2} \lambda_1^2 + \frac{3}{2}
	\lambda_2^2 \right) \frac{\xi}{4 \pi} \, , \\
\delta \lambda_{2,1l} & = & \left( \lambda_1 \lambda_2 + 2 \lambda_2^2
	\right) \frac{\xi}{4 \pi} \, , \\
Z_{\phi,1l}-1 & = & 0 \, , \\
Z_{m,1l} - 1 & = & \left( \frac{1}{2} \lambda_1 + \frac{1}{2} \lambda_2
	\right) \frac{\xi}{4 \pi} \, ,\\
\delta m^2_{1l} & = & - \left( \frac{1}{2} \lambda_1 
	+ \frac{1}{2} \lambda_2 \right) \frac{\Sigma}{4 \pi} \, , \\
\delta \< \phi^2 \>_{1l} & = & 2 \frac{\Sigma}{4 \pi} - 2 m^2
	\frac{\xi}{4 \pi} \, .
\ea
Note that, if $\lambda_1 = \lambda_2,$ then $\delta \lambda_1 = \delta
\lambda_2$; and similarly if $\lambda_1 = 3
\lambda_2$, then $\delta \lambda_1 = 3 \delta \lambda_2$.  Therefore the
decoupling and $O(2)$ symmetric versions of the theory are preserved
under renormalization.

\begin{figure}[t]
\centerline{\epsfxsize=12.5cm\epsfbox{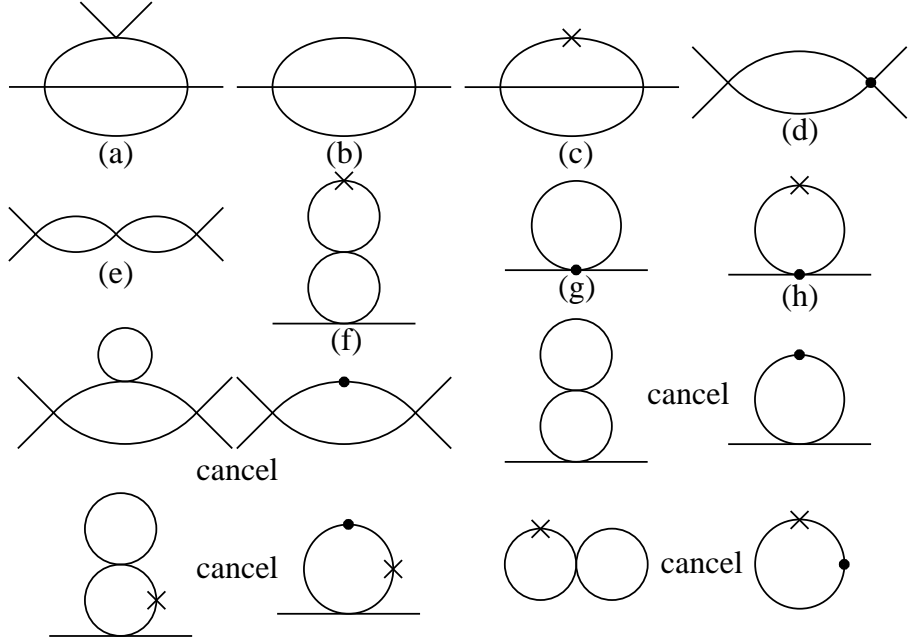}}
\caption{\label{two_loop_graphs} All required two loop graphs and one
loop graphs with one loop counterterm insertions, shown as heavy dots on
lines (mass counterterms) or at vertices (coupling counterterms).   The
last eight graphs cancel in pairs.  Diagrams (a), (d), and (e) are not
separately IR convergent; diagram (d) must be distributed between the
other two to produce IR convergent integrals.}
\end{figure}

It makes no sense to carry the matching to two loops unless we
use the improved lattice Laplacian, as $O(a^2)$ errors would already
appear at two loop level.  The two loop results require several more
graphs and the inclusions in one loop graphs of the one loop mass and
coupling counterterms, see \fig\ref{two_loop_graphs}.  Three more
integrals are needed, and their evaluation is detailed in the appendix.
The complete two loop renormalization is given by
\ba
\delta \lambda_1 - \delta \lambda_{1,1l} & = & - \left( 3 \lambda_1^3 
	+ 3 \lambda_1 \lambda_2^2 + 6 \lambda_2^3 \right) \frac{C_1}{16
	\pi^2} 
	+ \left( \frac{3}{4} \lambda_1^3 + \frac{9}{4} \lambda_1
	\lambda_2^2 \right) \left( \frac{\xi}{4 \pi} \right)^2  , \\
\delta \lambda_2 - \delta \lambda_{2,1l} & = & - \left( \lambda_1^2
	\lambda_2 + 6 \lambda_1 \lambda_2^2 + 5 \lambda_2^3 \right) 
	\frac{C_1}{16 \pi^2} 
	+ \left( \frac{3}{4} \lambda_1^2 \lambda_2 + \frac{9}{4} 
	\lambda_2^3 \right) \left( \frac{\xi}{4 \pi} \right)^2  , \\
Z_\phi - 1 & = & \left( \frac{1}{6} \lambda_1^2 + \frac{1}{2}
	\lambda_2^2 \right) \frac{C_2}{16 \pi^2} \, , \\
Z_m - Z_{m,1l} & = & \frac{1}{4} (\lambda_1 + \lambda_2)^2 \left
	( \frac{\xi}{4 \pi} \right)^2 - \left( \frac{1}{2} \lambda_1^2 +
	\frac{3}{2} \lambda_2^2 \right) \left( \frac{C_1}{16 \pi^2}
	\right) \, , \\
\delta m^2 - \delta m^2_{1l} & = & \left( \frac{1}{6} \lambda_1^2 + 
	\frac{1}{2} \lambda_2^2 \right) \frac{1}{16 \pi^2} \left( 
	\ln\frac{6}{a\mu} + C_3 \right) \nonumber \\ & & 
	- \left( \frac{\delta
	\lambda_1}{2} + \frac{\delta \lambda_2}{2} \right)
	\frac{\Sigma}{4 \pi} + 	(1-Z_{m,1l}) \delta m^2_{1l} \\
\hspace{-0.2in} \delta \< \phi^2 \> - \delta \< \phi^2 \>_{1l} & = & 
	\left( \lambda_1 + \lambda_2 \right) 
	\frac{\xi \Sigma}{16 \pi^2} \, .
\ea
Three new numerical constants appear here; their numerical values 
are $C_1 = .0550612,$ $C_2 =
.0334416,$ and $C_3 = -.86147916$.  $\delta m^2$ and $\delta \< \phi^2
\>$ can be defined when using Eq.~(\ref{unimp_Laplace}), in which case 
$C_{3,{\rm U}} = .08848010$; but the other improvements do not make
sense and should not be applied if that Laplacian is used.
We can again check that the form of $\delta \lambda_1$ and $\delta
\lambda_2$ is consistent at the two special values $\lambda_1 =
\lambda_2$ and $\lambda_1 = 3 \lambda_2$.

It is possible in principle to extend the improvement scheme to
$O(a^3)$, by making a three loop calculation.  However, at this order it
is necessary to include mixing of different dimensions of operator
insertions and to include counterterms for radiatively induced high
dimension operators in the Lagrangian.  The calculation of the graphs
also becomes significantly more challenging.  The improvement presented
here is sufficient for our purposes; the leading lattice spacing errors
in measurables related to the strength of the phase transition 
now first appear at $O(a^3)$.

\subsection{Two loop graphs}

Here we detail the calculation of $C_1$, $C_2$, and $C_3$.

We begin with the two loop vertex correction, Figure
\ref{two_loop_graphs} graph (a).  The required integral, including the
appropriate amount of the one loop counterterm graph (d), is
\be
\frac{C_1}{16 \pi^2} = 
	\int_{k,BZ} \frac{1}{(\tilde{k}^2)^2} \left\{ \int_{l,BZ} 
	\frac{1}{\tilde{l}^2 \widetilde{(k+l)}^2} - \frac{\xi}{4 \pi} 
	\right\} - \int_{k,\Re^3} \frac{1}{k^4} \int_{l,\Re^3} 
	\frac{1}{l^2 (k+l)^2} \, .
\ee
Here we use the shorthand in which the integration limit also lists
which variable is being integrated over; so $\int_{k,BZ}$ means
$\int_{[-\pi,\pi]^3} (d^3 k/(2 \pi)^3)$.
The first step is to evaluate the inner continuum integral, which can be
done by standard Feynman parameter methods;
\be
\int_{l,\Re^3} \frac{1}{l^2 (k+l)^2} = \int_0^1 d\alpha \frac{1}{2
\pi^2} \int_0^\infty \frac{l^2 dl}{(l^2 + \alpha(1-\alpha)k^2)^2} = 
\frac{1}{8k} \, .
\ee
Then we re-arrange the original integral into three parts,
\be
\int_{k,BZ} \frac{1}{(\tilde{k}^2)^2} \left\{ \int_{l,BZ}
	\frac{1}{\tilde{l}^2 \widetilde{(k+l)}^2} - \frac{1}{8k} -
	\frac{\xi}{4 \pi} \right\} + \int_{k,BZ} \frac{1}{8k} \left( 
	\frac{1}{(\tilde{k}^2)^2} - \frac{1}{k^4} \right) 
	- \int_{k,\Re^3-BZ} \frac{1}{8 k^5} \, .
\ee
The first integral is IR well behaved because the two counterterms
cancel the $l$ integral up to a $k^2$ correction, which in the small $k$
limit is $.0125438 k^2 / 4\pi$.  The integrals can be performed by
quadratures using adaptive mesh refinement techniques and Richardson
extrapolation. The first integral gives $.0360003/16 \pi^2$ 
and the second gives $.054568958/16 \pi^2$.
The last integral, over $\Re^3-BZ$, can
be re-arranged into
\be
-\frac{3}{16 \pi^5} \int_0^1 dx \int_0^1 dy \frac{1}{(1+x^2+y^2)^{5/2}}
	= \frac{-.035507296027 \ldots}{16 \pi^2} \, .
\la{easy_int}
\ee
These sum to give $C_1 = .0550612$.

Besides this graph there is graph (e), which gives
\be
\left( \int_{k,BZ} \frac{1}{(\tilde{k}^2)^2} \right)^2 - 
	\left( \int_{k,\Re^3} \frac{1}{k^4} \right)^2 \, ,
\ee
which is {\em not} IR convergent; however, including -2 times the
counterterm diagram (d),
\be
-2 \left( \int_{k,BZ} \frac{1}{(\tilde{k}^2)^2} \right)
	\left(  \int_{k,BZ} \frac{1}{(\tilde{k}^2)^2} - 
	 \int_{k,\Re^3} \frac{1}{k^4} \right) \, ,
\ee
gives $-(\xi/4\pi)^2$; no new integrals are required.  It is a
nontrivial check on the calculation that the sum of the coefficients
arising from diagrams (a) and (e) precisely absorb diagram (d).

The next integral is the $O(p^2)$ contribution from the setting sun
diagram (b), 
\ba
\frac{C_2}{16 \pi^2} & = & \lim_{p\rightarrow 0} \frac{1}{p^2}
	\Bigg\{ \int_{k,BZ} \left[ \left( 
	\frac{1}{\widetilde{(k+p)}^2} - \frac{1}{\tilde{k}^2} \right) 
	\left( \int_{l,BZ} \frac{1}{\tilde{l}^2 \widetilde{(k+l)}^2} 
	\right) \right] \nonumber \\ & & \qquad \qquad 
	- \int_{k,\Re^3} \left[ \left( \frac{1}{(k+p)^2}
	-\frac{1}{k^2} \right) \left( 
	\int_{l,\Re^3} \frac{1}{l^2(k+l)^2} \right) \right] \Bigg\} \, .
\la{messy1}
\ea
The first trick is to note that 
\be
\int_{k,BZ} \left( \frac{1}{\widetilde{(k+p)}^2} 
	- \frac{1}{\tilde{k}^2} \right) = 0
\ee
just by shifting the integration variable for the first term; so we may
add $-\xi/4\pi$ to the term in the second parenthesis of the first line
of Eq.~(\ref{messy1}).  This prevents IR divergences
in what follows, so we are free to expand $1/\widetilde{(k+p)}^2$ to
second order in $p$; after averaging over directions for $p$, we find 
\be
\frac{1}{\widetilde{(k+p)}^2} - \frac{1}{\tilde{k}^2} = p^2 \left[
	\frac{\frac{1}{3} \sum_i \left( \frac{8 \sin k_i - 
	\sin 2 k_i}{3} \right)^2 }{(\tilde{k}^2)^3} - \frac{ \frac{1}{3} 
	\sum_i \left( \frac{4 \cos k_i - \cos 2k_i}{3} \right)}
	{(\tilde{k}^2)^2} \right] \equiv p^2 {\cal M}(k) \, .
\ee
The equivalent expression in the continuum case is $(p^2/3)/k^4$.
Re-arranging the terms a little, we can write
\ba
\frac{C_2}{16 \pi^2} & = & -\frac{1}{24} \int_{k,\Re^3-BZ} \frac{1}{k^5}
	+ \int_{k,BZ} \frac{1}{8k} \left( {\cal M}(k) -
	\frac{(1/3)}{k^4} \right) \nonumber \\ & & 
	+ \int_{k,BZ} {\cal M}(k) 
	\left[ \int_{l,BZ} \frac{1}{\tilde{l}^2 \widetilde{(k+l)}^2}
	- \frac{1}{8k} - \frac{\xi}{4 \pi} \right] \, .
\ea
The first integral is (1/3) of Eq.~(\ref{easy_int}).
The second gives $.0310757695/16 \pi^2$ and the
last gives $.0142016/16 \pi^2$, so $C_2 = .0334416$.

Next we must compute the $O(p^0)$ part of the setting sun diagram.
The continuum diagram is log IR and UV divergent, while the lattice
diagram is only log IR divergent.  It is convenient to IR regulate both
by introducing a mass on one line.  In this case the continuum integral
can be performed in $\overline{\rm MS}$, leaving a lattice integral
minus an analytically determined counterterm \cite{FKRS,LaineRajantie}.
Choosing to separate the renormalization dependence along with the same
finite constant as in the previous literature \cite{FKRS,LaineRajantie},
the constant $C_3$ is given by
\be
\frac{C_3}{16 \pi^2} = \lim_{m \rightarrow 0} \left\{ \int_{k,BZ} 
	\frac{1}{\tilde{k}^2 + m^2} \int_{l,BZ}
	\frac{1}{\tilde{l}^2 \widetilde{(k+l)}^2}
	- \frac{1}{16 \pi^2} \left[ \frac{1}{2} 
	+ \ln \frac{6}{m} \right] \right\} \, .
\ee
The annoying feature of this expression is the logarithm.  To remove it,
we add and subtract $1/8k$ to the integral over $l$.  The integral
\be
\int_{k,BZ} \frac{1}{\tilde{k}^2 + m^2} \left[ 
	\int_{l,BZ} \frac{1}{\tilde{l}^2 \widetilde{(k+l)}^2} - 
	\frac{1}{8k} \right]
\ee
is IR convergent and the $m \rightarrow 0$ limit may be taken
immediately.  It evaluates to $-.06858432/16 \pi^2$, unless we use the
unimproved lattice Laplacian, in which case it is $.60953343 / 16
\pi^2$.  We re-arrange the remaining terms to be
\be
\int_{k,BZ} \left( \frac{1}{\tilde{k}^2 + m^2} - \frac{1}{k^2 + m^2} 
	\right) \frac{1}{8k} + \int_{k,BZ} \frac{1}{8k(k^2+m^2)} - 
	\frac{1}{16 \pi^2} \left[ \frac{1}{2} + \frac{6}{m} \right] 
	\, .
\ee
Again, for the first integral the $m \rightarrow 0$ limit may be taken
immediately, and the numerical value is $.161799607 / 16 \pi^2$, or
$.43364112015 / 16 \pi^2$ if we use the unimproved Laplacian.  For the
last integral, we cut the integration region into the ball of radius
$\pi$ and the region within the Brillouin zone but outside the ball:
\be
\int_{k,BZ}\frac{1}{8k(k^2+m^2)} = \frac{1}{2 \pi^2} \int_0^\pi 
	\frac{k^2 dk}{8k(k^2+m^2)} + \int_{k,BZ} \frac{1}{8k(k^2+m^2)} 
	\Theta(|k|-\pi) \, .
\ee
The former may be promptly integrated to give $\ln(\pi/m)/16 \pi^2$ plus
terms power suppressed in $m$; when added to $(-1/16 \pi^2) (\ln(6/m) +
1/2)$ this cancels the $\ln(m)$, leaving $(1/16\pi^2)(\ln(\pi/6)
- 1/2)$.  The final integral has had the small $k$ part of the
integration range removed, so again the $m \rightarrow 0$ limit may be
taken.  It can then be reduced to
\be
\frac{1}{16 \pi^2} \int \frac{d\Omega}{4 \pi} \ln(R({\rm cube})-R({\rm
	ball})) = \frac{1}{16 \pi^2} \frac{12}{\pi} \int_0^{\pi/4}
	d\phi \int_0^{{\rm arctan}( \sec \phi)} \sin(\theta) d\theta 
	\ln(\sec(\theta)) 
\ee
which numerically equals $.19233513195/16 \pi^2$.  Note that at no point
have we had to deal numerically with an integral which is log divergent
in $m$, or which still contains $m$ at all.

Combining terms gives $C_3 = -.86147916$, unless we use the unimproved
lattice Laplacian, in which case it is $C_3 = .08848010$.  In the
notation of \cite{FKRS,LaineRajantie}, $C_3$ is called $\zeta$.  Note
that, unlike $C_1$ and $C_2$, $C_3$ is relatively large.  Similarly,
$\xi$ is small but $\Sigma$ is large.  This means that the radiative
$O(a)$ and $O(a^2)$ corrections to quantities which do not renormalize
in the continuum are all small, but the corrections to the mass are
larger.  The size of $C_3$ also depends on a somewhat arbitrary choice
to make it accompany $\ln (6/a\mu)$.


\end{document}